\documentclass[a4paper,11pt]{article}
\usepackage[natbib=true, style=numeric,sorting=none]{biblatex} 
\usepackage{epsfig,amsmath,amssymb}
\usepackage{graphicx}
\usepackage{xcolor}
\usepackage[colorlinks=true,allcolors=blue]{hyperref}
\usepackage{subfigure}
\usepackage{float}
\usepackage{overpic}
\usepackage[normalem]{ulem}
\usepackage{simpler-wick}
\usepackage[utf8]{inputenc}

\def\be{\begin{equation}}
\def\ee{\end{equation}}
\def \ba {\begin{array}}
\def \ea {\end{array}}
\def \bea{\begin{eqnarray}}
\def \eea{\end{eqnarray}}
\def \a {\alpha}
\def \b {\beta}
\def \g {\gamma}

\def \d {\delta}

\def \e {\epsilon}

\def \m {\mu}
\def \n {\nu}

\def \l {\lambda}

\def \s {\sigma}
\def \S {\Sigma}
\def \r {\rho}

\def \t {\tau}

\def \p {\partial}

\def \nn {\nonumber}

\def \mb {\mathbf}

\def\cA{{\cal A}} \def\cB{{\cal B}} \def\cC{{\cal C}}
\def\cD{{\cal D}} \def\cE{{\cal E}} \def\cF{{\cal F}}
 \def\cH{{\cal H}} 
  \def\cL{{\cal L}}
\def\cM{{\cal M}}  \def\cO{{\cal O}}
  \def\cR{{\cal R}}
\def\cS{{\cal S}}  
 \def\cW{{\cal W}}

\def\hs{\hspace}
\def\Tr {\mbox{Tr}}

\numberwithin{equation}{section}

\newcommand{\cb}[1]{\textcolor{purple}{ #1-CB}}

\begin{document}

\title{Reflected Entropy in  AdS$_3$/WCFT}
 
 \author{
Bin Chen$^{1,2,3}$, Yuefeng Liu$^1$ and Boyang Yu$^2$}
\date{}

\maketitle

\vspace{-10mm}

\begin{center}
{\it

$^1$Department of Physics, Peking University, No.5 Yiheyuan Rd, Beijing
100871, P.R. China\\\vspace{1mm}

$^2$Center for High Energy Physics, Peking University,
No.5 Yiheyuan Rd, Beijing 100871, P. R. China\\\vspace{1mm}

$^3$ Collaborative Innovation Center of Quantum Matter,
No.5 Yiheyuan Rd, Beijing 100871, P. R. China\\\vspace{1mm}

}
\end{center}

\vspace{8mm}

\begin{abstract}
Reflected entropy is a newly proposed notion in quantum information. It has important implications in holography. In this work, we study the reflected entropy in the framework of the AdS$_3$/WCFT correspondence. We determine the scaling dimensions and charges of various twist operators in non-Abelian orbifold WCFT by generalizing the uniformization  map and taking into account of the charge conservation. This allows us to compute the reflected entropy, logarithmic negativity and odd entropy for two disjoint intervals in holographic WCFT. We find that the reflected entropy can be related holographically to the pre-entanglement wedge cross-section, which is given by the minimal distance between the  benches in two swing surfaces.   
\end{abstract}
\maketitle
\newpage
\tableofcontents 
\newpage

\section{Introduction}

The quantum entanglement may play a fundamental role in the emergence of spacetime. In the past fifteen years, there have been a lot of works on the holographic entanglement entropy and its implications, since the seminal works in \cite{Ryu:2006bv,Ryu:2006ef}(see \cite{Rangamani:2016dms} for a nice review). It is widely believed that the entanglements in  holographic states encode the information of dual (quantum) gravity. Various hard-core problems in AdS/CFT have been addressed from entanglement point of view. 

  The entanglement entropy is defined as the von Neumann entropy of the reduced density matrix, and it smooths out the details of the reduced density matrix. As a result, the entanglement entropy does not capture the full entanglement structure in a quantum state. 
  In particular, for the mixed state, it is not a proper entanglement measure: it captures classical (thermal) correlations as well as purely quantum ones. 
  
  One can purify a mixed state  by combining it with an auxiliary system such that the combined system is a pure state. There are infinite number of ways to purify a mixed state. One may define the entanglement of purification (EoP) to be the minimal entanglement in all purification. Even though it is difficult to find EoP in the quantum systems, it was proposed in \cite{Takayanagi:2017knl,Nguyen:2017yqw} that  
 EoP could be holographically dual to the entanglement wedge cross section.
 
 The entanglement wedge is defined to be the bulk causal domain of a co-dimension one surface interpolating between the boundary subregion and its holographic entanglement surface. It plays an important role in studying the bulk reconstruction and the subregion/state duality\cite{Czech:2012bh}. The entanglement wedge cross-section (EWCS) is the minimal area surface in the bulk that divides the entanglement wedge of two disconnected subregions.   Remarkably, besides the EoP,  several quantum information measures have been proposed to be related to EWCS, including the logarithmic entanglement negativity\cite{Peres:1996dw,Kudler-Flam:2018qjo}, odd entropy\cite{Tamaoka:2018ned} and  reflected entropy\cite{Dutta:2019gen}.


 The reflected entropy $S_{R}(A:B)$ is defined for a bipartite quantum system $A \cup B $ and a mixed state denoted by a density matrix $\rho_{AB}$ on the Hilbert space $\mathcal{H}=\mathcal{H}_{A}\otimes \mathcal{H}_{B}$. Analogous to the relationship between the thermofield double state and thermal density matrix, one can define the canonical purification of this density matrix in a doubled Hilbert space $\mathcal{H}\otimes \mathcal{H}' = \mathcal{H}_{A} \otimes  \mathcal{H}_{B} \otimes \mathcal{H}_{A'} \otimes \mathcal{H}_{B'}$. Then the reflected entropy is defined as the entanglement entropy or the von Neumann entropy of the density matrix $\rho_{AA'}$:
\be S_{R}(A:B)=S(AA')_{\sqrt{\rho_{AB}}}= S(BB')_{\sqrt{\rho_{AB}}}=S_{vN}(\rho_{AA'})  \ee
where $\rho_{AA'}=\Tr_{BB'}|\sqrt{\rho_{AB}}\rangle \langle \sqrt{\rho_{AB}}|$ is the reduced density matrix after tracing over $\mathcal{H}_{B}\otimes \mathcal{H}_{B'}$. For the recent studies on the reflected entropy in quantum information and quantum field theories, see \cite{Bueno:2020vnx,Berthiere:2020ihq,Kudler-Flam:2020xqu,Bueno:2020fle,Camargo:2021aiq,Hayden:2021gno,Akers:2021pvd,Akers:2022max,Basu:2022nds,Basak:2022cjs,BasakKumar:2022stg}. 
  For holographic CFT, it was proposed \cite{Dutta:2019gen} that the reflected entropy  is  related to  the entanglement wedge cross-section (EWCS) 
 \be 
S_{R}(A:B)=2\frac{\mbox{Area}(EWCS)}{4G_N} +\mbox{(subleading terms)}, \label{REWCS}
\ee 
where $G_N$ is the Newton's constant and the subleading terms include the quantum corrections in the small $G_N$ expansion.  
 Another remarkable point is that the reflected entropy can be useful in understanding the requirement of tripartite entanglement in holographic states\cite{Akers:2019gcv,Hayden:2021gno}.



In this work, we would like to study the reflected entropy in two-dimensional (2D) holographic warped conformal field theory (WCFT) and  its bulk dual within the framework of AdS$_3$/WCFT. The 2D warped CFT is a field theory whose symmetry is generated by  Virasoro-Kac-Moody algebra\cite{Hofman:2011zj,Detournay:2012pc}. The holographic WCFT could be dual to semiclassical  AdS$_3$ gravity with the Comp\`ere-Song-Strominger (CSS) boundary condition\cite{Compere:2013bya} or the 3D topological massive gravity in warped AdS spacetime. The resulting AdS$_3$/WCFT and WAdS$_3$/WCFT correspondences provide nontrivial windows to study the holographic duality beyond the AdS/CFT correspondence. We focus on the AdS$_3$/WCFT case in this work.
The AdS$_3$/WCFT correspondence has been studied from various points of view in \cite{Compere:2013aya,Song:2016gtd,Song:2017czq,Castro:2017mfj,Apolo:2018eky,Wen:2018mev,Apolo:2018oqv,Song:2019txa,Chen:2019xpb,Lin:2019dji,Bao:2019zqc,Chen:2020juc,Apolo:2020qjm,Caceres:2020jcn}. 

In the study of holographic entanglement entropy in AdS/WCFT, there appears some novel features. For a single interval, one may use the generalized Rindler method to find the swing surface, whose area gives the entanglement entropy\cite{Castro:2015csg,Song:2016gtd}. Quite recently, it was proposed in \cite{Apolo:2020bld,Apolo:2020qjm} to use the modular Hamiltonian to find the swing surface. In contrast to the usual AdS/CFT correspondence, the definition of the entanglement wedge in AdS$_3$/WCFT is subtle, since the homology surface interpolating between boundary interval and the bench in the swing is not always well-defined. Nevertheless,  one can still define pre-entanglement wedge and moreover the pre-entanglement wedge cross section (pre-EWCS) of two disjoint intervals. 

As the first step, we need to compute the reflected entropy in a WCFT. The computation turns out to be  challenging. The main difficulty is how to  determine  the dimensions and charges of the twist operators in the non-Abelian orbifold. We need to generalize the uniformization map proposed in the Abelian orbifold\cite{Chen:2019xpb}  to the case at hand. Moreover we choose the monodromy conditions on the twist fields in a consistent way such that the charge conservation is kept. As a result, not only can we  compute the reflected entropy, but also other two information quantities, the logarithmic negativity and odd entropy. For the holographic WCFT we will show that the conjectured relation \eqref{REWCS} between the reflect entropy and EWCS is still true even in the AdS$_3$/WCFT correspondence.



 The remaining parts of the paper are organized as follows. In section 2, we review the properties of the reflected entropy, its computations in 2D CFT and its implication in the AdS$_3$/CFT$_2$ correspondence. In section 3, in order to compute the reflected entropy in a WCFT, we show how to determine the dimensions and the charges of various twist operators in a non-Abelian orbifold, by using a generalized uniformization map and charge conservation. In particular we calculate the reflected entropy of two disjoint intervals in holographic WCFT.  In section 4, we discuss the pre-entanglement wedge cross section in AdS$_3$ gravity with CSS boundary condition, and its relation with the reflected entropy in the dual field theory. In section 5, with the general discussions on the twist operators in section 3, we compute two other information quantities, logarithmic negativity and odd entropy, in the holographic WCFT. We end with conclusion and some discussions in section 6. 

 
 
 
 
\section{Reflected Entropy and Holographic 2D CFT}

In this section we would like to give a concise review on the reflected entropy $S_{R}(A:B)$ of two disjoint intervals $A,B$ and its holographic description in the standard AdS$_3$/CFT$_2$ correspondence. The review is mainly based on \cite{Dutta:2019gen,Akers:2021pvd}.

\subsection{Reflected entropy: finite dimensional case} 

It is easier to start with the definition of reflected entropy in finite dimensional Hilbert space. Suppose that we have a bipartite quantum system $A \cup B $ and a mixed state denoted by a density matrix $\rho_{AB}$ on the Hilbert space $\mathcal{H}=\mathcal{H}_{A}\otimes \mathcal{H}_{B}$. Analogous to the relationship between the thermofield double state and thermal density matrix, we can define the canonical purification of this density matrix in a doubled Hilbert space $\mathcal{H}\otimes \mathcal{H}' = \mathcal{H}_{A} \otimes  \mathcal{H}_{B} \otimes \mathcal{H}_{A'} \otimes \mathcal{H}_{B'}$. Actually there is a natural mapping between the space of linear operators acting on $\mathcal{H}$ and the spaces of states on this doubled Hilbert space $\mathcal{H}\otimes \mathcal{H}'$ with the inner product
 \be \langle \rho| \sigma \rangle_{ABA'B'}=\Tr_{AB}(\rho^{\dagger}\sigma).  \ee
Thus, the operator $\sqrt{\rho_{AB}}$ is mapped to a state $|\sqrt{\rho_{AB}} \rangle_{ABA'B'} $ which is the canonical purification of $\rho_{AB}$. It is not hard to show that the original density matrix $\rho_{AB}$ can be recovered by tracing out the subregions $A' \cup B'$
\be \Tr_{A'B'}|\sqrt{\rho_{AB}}\rangle \langle \sqrt{\rho_{AB}}|= \rho_{AB}.\ee
So the above construction does represent a genuine purification. The reflected entropy $ S_R(A:B) $ is defined as the entanglement entropy or the von Neumann entropy of the density matrix on the subregion $A\cup A'$ in the pure state $|\sqrt{\rho_{AB}}\rangle$
\be S_{R}(A:B) \equiv S(AA')_{\sqrt{\rho_{AB}}}= S(BB')_{\sqrt{\rho_{AB}}}=S_{vN}(\rho_{AA'}),  \ee
where $\rho_{AA'}=\Tr_{BB'}|\sqrt{\rho_{AB}}\rangle \langle \sqrt{\rho_{AB}}|$ denotes the reduced density matrix after tracing over $\mathcal{H}_{B}\otimes \mathcal{H}_{B'} $.  One remarkable properties of reflected entropy is that it is bounded by the mutual information $I(A:B)$ and original entanglement entropies $S(A)$, $S(B)$, 
\be
I(A:B)\leq S_R(A:B)\leq 2\mbox{min}(S(A),S(B)).
\ee
For more properties of the reflected entropy, please refer to \cite{Dutta:2019gen}.

\subsection{Reflected entropy in conformal field theory}  

Although the above definition is adapted to the finite dimensional Hilbert space, we can directly generalize the definition of $S_{R}(A:B)$ to the continuous field theories with an infinite dimensional Hilbert space, where the situation is very similar to the case of defining the entanglement entropy in field theory. By using the replica trick, we may first define the R\'enyi reflected entropy $S_R^{(n)}(A:B)$, and then taking the limit that the replica number $n$ goes to $1$ to read the reflected entropy, under the assumption that the analytical continuation in $n$ is feasible. In general, the R\'enyi reflected entropy is given by the partition functions of original field theory on a replicated manifolds, which may have nontrivial topology and geometry. In 2D CFT, due to the infinite dimensional conformal symmetry and the fact that the twist operators inducing the identification of the fields in doing the replication are local, the partition function can be transformed into the multi-point correlation functions of the twist operators in an orbifold CFT. 

However, unlike the case of entanglement entropy, here we need to do double replication by making $n \times m$ copies of the original theories in order to both represent the purified state $|\sqrt{\rho_{AB}} \rangle_{ABA'B'} $ and evaluate its Renyi reflected entropy in a path integral language. The first replication replace the canonically purified state by its $m$ copies 
\be  |\psi_{m}\rangle=\frac{1}{\sqrt{\Tr\rho_{AB}^{m}}}|  \rho_{AB}^{m/2} \rangle, \ee
which is normalized. This state can be described by a path-integral when $m\in 2\mb{Z}^+$. The second replication is necessary to compute the $(m,n)$-R\'enyi reflected entropy for a positive integer $n$. We would like to compute 
\be  S^{(m,n)}_{R}(A:B)=\frac{1}{1-n} \log{\Tr\big( \rho_{AA'}^{(m)} \big)^{n}} \ee
where 
\be \rho_{AA'}^{(m)} = \Tr_{BB'} |\psi_{m}\rangle \langle \psi_{m}|= \frac{1}{\Tr \rho_{AB}^{m}} \Tr_{BB'}|\rho_{AB}^{m/2} \rangle  \langle \rho_{AB}^{m/2} |.\ee
Introducing $Z_{n,m}$ to represent the un-normalized partition functions 
\be Z_{n,m}=\Tr_{AA'}\big( \Tr_{BB'} |\rho_{AB}^{m/2} \rangle  \langle \rho_{AB}^{m/2} | \big)^{n}, \label{Znm}\ee
which satisfy $Z_{1,m}=\Tr\rho_{AB}^{m}$, then we find 
\be S^{(m,n)}_R(A:B)=\frac{1}{1-n} \log{\frac{Z_{n,m}}{(Z_{1,m})^{n}}}. \ee
Note that the partition function $Z_{n,m}$ can be evaluated by a path integral on a replicated manifold $\Sigma_{n,m}$ for $n\in \mb{Z}^{+}$ and $m \in 2 \mb{Z}^{+}$. Note that the number of replica sheets in the replicated manifold of $|\rho_{AB}^{m/2} \rangle$ is $m/2$ which should be a positive integer. Then by using the analytic continuation in $m$ and $n$, we can evaluate the reflected entropy by taking the limit: \be
S_R(A:B)=\lim_{n,m\to 1} S^{(m,n)}(A:B). \ee

Another way to obtain the partition function \eqref{Znm} is to compute the correlation function of the (generalized) twist operators, which induce the identifications between the fields at different replicas. For disjoint intervals $A$ and $B$ in general dimensions, we can rewrite and reinterpret the partition functions $Z_{n,m}$ in \eqref{Znm} as follows \cite{Dutta:2019gen} 
\be 
Z_{n,m}=\langle \S_{g_B}(B)\S_{g_A}(A)\rangle_{\mbox{CFT}^{\otimes nm}} \label{twistAB}
\ee
where $\S$'s are twist operators located at $\p A$ or $\p B$ in the product theory $\mbox{CFT}^{\otimes nm}$. The meaning of the twist operators needs some clarifications. Usually in higher dimensions ($D\geq 3$), the twist operators are truly non-local and hard to work with, while in two dimensions the twist operators can be seen as the local fields located at each endpoints. Moreover, if the original 2D field theory is conformal invariant, the twist operators in it are actually primary fields, and hence their correlation functions can be discussed using various analytical CFT techniques.  However, there is an additional complication when actually calculating the reflected entropy, which is related to the fact that we choose not to gauge the product theory $\mbox{CFT}^{\otimes nm}$. We would clarify this point more later. Now we are taking double replication and thus have $n\times m$ copies of original CFT, each living on a flat Euclidean spacetime. There is a symmetric permutation group $S_{nm}$ related to these replicas. We can label each replica with $(\a,\b)\in (\mb{Z}_{n}^{+},\mb{Z}_{m}^{+})$, and represent the actions of the special group elements of $S_{nm}$ which are needed in the definition of $\S$ operators by
\be 
\t_m^{(\n)}(\a,\b)=(\a,\b+\d_{\a,\n}),\hs{3ex}\t_n^{(\m)}(\a,\b)=(\a+\d_{\b,\m},\b). \label{labelnm} \ee 
The first action $\t_m^{(\n)}(\a,\b)$ defines the cyclic permutation on the fixed $n$-replica along $m$ direction, while the second one $\t_n^{(\m)}(\a,\b)$ defines the cyclic permutation on the fixed $m$-replica along $n$ direction. The full $m$-cyclic permutation is simply the product of all $\t_m^{(\n)}(\a,\b)$, i.e., $g_m=\prod_{\n=0}^{n-1}\t_m^{(\n)}$, and similarly for $g_n$. From the Euclidean path integral in computing the partition functions $Z_{nm}$, we can read the group elements which identify the fields in different replicas in an appropriate way:
\bea 
\p A:&&g_A=(\t_n^{(0)})^{-1}\t_n^{(m/2)}g_m, \label{gaga}\\
\p B:&&g_B=g_m. \label{gagb}
\eea 
These two group elements help us to define  co-dimension-two twist operators $\S(A)$ at $\p A$ and $\S(B)$ at $\p B$ in \eqref{twistAB}.  Note that the two group elements $g_{A}$ and $g_{B}$ defined in (\ref{gaga},\ref{gagb}) are conjugate to each other. For more complete discussions and properties about the group elements defined here, see \cite{Dutta:2019gen}.

\begin{figure}
    \centering
    \includegraphics[scale=0.6]{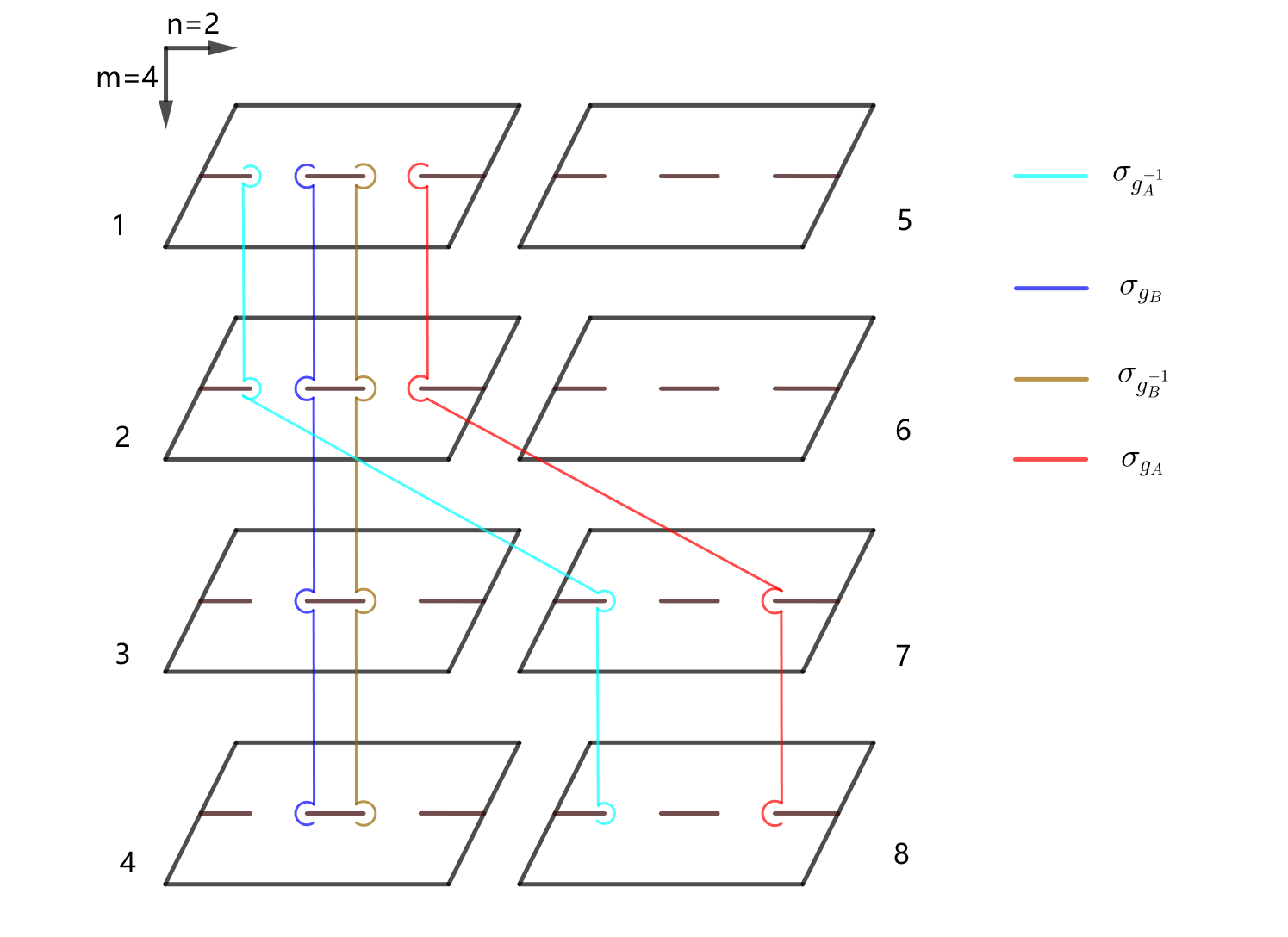} 
    \caption{(Half of) different monodromy conditions or field identifications related to the four twist operators $\sigma_{g_{A}}$, $\sigma_{g_{A}^{-1}}$, $\sigma_{g_{B}}$ and $\sigma_{g_{B}^{-1}}$ in the case $m=4,n=2$. The twist operator $\sigma_{g_{A}}$ can be equivalently represented in another useful notation $\sigma_{(1278)(3456)}$ for this particular picture, where $(1278)$ denotes the cyclic permutation of these four numbers. Here to keep the picture tidy and clean, we only show  half of the identifications $(1278)$. }
    \label{fig:monodromy3}
\end{figure}

Let us specialize to 2D CFT and still choose two disjoint intervals $A=(x_{1},x_{2})$ and $B=(x_{3},x_{4})$ on the same time-slice with $x_{1}<x_{2}<x_{3}<x_{4}$. Here the two  twist operators $\S(A)$, $\S(B)$ become four (quasi) local twist operators $\s_{g_{A}}$, $\s_{g_{A}^{-1}}$, $\s_{g_{B}}$ and $\s_{g_{B}^{-1}}$ located at each endpoints separately as mentioned above. Then the partition functions \eqref{Znm} is captured by the following four-point function 
\begin{align}
    Z_{n,m}&= \langle \sigma_{{g}_A}(x_1) \sigma_{{g}^{-1}_A}(x_2) \sigma_{{g}_B}(x_3) \sigma_{{g}^{-1}_B}(x_4) \rangle _{CFT^{\otimes mn}} , \\
    Z_{1,m}&= \langle \sigma_{{g}_m}(x_1) \sigma_{{g}^{-1}_m}(x_2) \sigma_{{g}_m}(x_3) \sigma_{{g}^{-1}_m}(x_4) \rangle _{CFT^{\otimes m}} . 
\end{align}
The twist operators $\sigma_{g_{A}}$, $\sigma_{g_{A}^{-1}}$, $\sigma_{g_{B}}$ and $\sigma_{g_{B}^{-1}}$ are defined by the specific boundary conditions on the replica sheets, and the twist operators $\sigma_{g_{m}}$ and $\sigma_{g_{m}^{-1}}$ represent the usual cyclic and anti-cyclic permutations among $m$ replica sheets, which can be read from the $n\to 1$ limit of $\sigma_{g_{A}}$, $\sigma_{g_{B}}$ operators: 
\be  \sigma_{g_{m}}=\lim_{n\to 1}\sigma_{g_{A}}=\lim_{n\to 1}\sigma_{g_{B}}. \label{sigmam} \ee
In Figure \ref{fig:monodromy3}, we show explicitly the monodromy conditions or field identifications of various twist operators in the case $m=4, n=2$. It can be interpreted as a path integral representation of the partition function $ Z_{2,4}=\Tr_{\cH_{A} \cH_{A'}} \left( \Tr_{\cH_{A} \cH_{A'} } | \rho_{AB}^{2} \rangle \langle \rho_{AB}^{2} | \right)^{2} $  with replicated $ 2\times 4$ sheets. 

There are a few remarkable features in the calculation of reflected entropy $S_{R}(A:B)$ of two disjoint intervals, compared with the ones of entanglement entropy or mutual information:  
\begin{itemize}
    \item In calculating the mutual information, the twist operators can be taken as the primary operators in an Abelian orbifold CFT $\mbox{CFT}^{\otimes}/ \mb{Z}^{n}$, while loosely speaking for the reflected entropy the twisted operators are in a non-Abelian orbifold CFT. The twist operators induce the field identifications in different replicas. In  the Abelian case, the field identifications are simply 
    \be \phi_{j}(w+(z-w)e^{i\theta}) \sigma_{n}(z,\bar{z})\to \phi_{j+1}(z) \sigma_{n}(z,\bar{z}),\quad \theta \in (0,2\pi),\quad j \in [1,n] \in \mb{Z}^{+},  \ee
    while in the non-Abelian case,  the field identifications may take  more involved forms
    \be \phi_{j}(w+(z-w)e^{i\theta}) \sigma_{p}(z,\bar{z})\to \phi_{p(j)}(z) \sigma_{p}(z,\bar{z}),\quad \theta \in (0,2\pi),\quad j \in [1,n] \in \mb{Z}^{+} \ee 
where $p\in S_{n}$ is an arbitrary group element in permutation group $S_n$ and $p(j)$ is the image of number $j$ under the group element $p$. A crucial point is that the permutation group element $p$ can have complicated multi-cycle structures. For example in Figure \ref{fig:monodromy3} the operator $\s_{g_{B}}$ can be rewritten in the form $\s_{(1234)(5678)}$ with $(1234)$ representing cyclic permutation of $(1,2,3,4)$. In any case, the holomorphic weight of the twist operator is
\be   h_{p}=\sum_{j=1}^{m} h_{p_{j    }}=\frac{c}{24}\sum_{j=1}^{m}(p_{j}-\frac{1}{p_{j}})  \label{nonabelian} \ee
which is simply the summation of the holomorphic weights of the ``twist" operators in separated cycles. This is closely related to the locality of stress tensor $T_{\mu\nu}$. Another important features is that the OPE of these arbitrary twist operators must respect  the group symmetry of $S_{n}$.

    \item The twist operators $\sigma_{g_{A}}$, $\sigma_{g_{A}^{-1}}$, $\sigma_{g_{B}}$ and $\sigma_{g_{B}^{-1}}$ are defined in a product theory $\mbox{CFT}^{\otimes mn}$, without gauging the permutation group $S_{nm}$. 
    Due to the specific boundary conditions of the replica manifolds in the path integral representation, or more precisely, due to the multi-cycle structure of $g_{A}$, $g_{B}$ group elements in defining the relevant twist operators, gauging the  permutation group $S_{nm}$ would not only generate local operators, but also introduce some complicated non-local objects connecting the boundaries of the twist operators. Thus we choose to directly work in the tensor product theory $\mbox{CFT}^{\otimes nm}$, and at the same time be careful with various subtleties originating from this choice.  
    
    \item  One subtlety related to the product theory $\mbox{CFT}^{\otimes nm}$ is that unlike the true non-Abelian orbifold, where different twist operators are labelled by different group conjugacy classes, the operator spectrum of the this product theory contain much more ``new" twist operators in the ``new" twisted sectors, that would be identified as the same one after gauging the group $S_{n}$. Thus we should be careful about the evaluation of the OPE coefficients of the twist operators, due to the great richness of spectrum. Note that usually in the evaluation of entanglement entropy $S_{A} $ or mutual information $I(A:B)$ in holographic 2D CFT, the OPE coefficient in the expansion
    \be \s_{n} \Tilde{\s_{n}} \sim \mathrm{1}+...  \ee 
    is of order $1$, and thus this OPE coefficient do not play an important role  in the final answer of $S_{A}$ or $I(A:B)$. While in the reflected entropy case, as explicitly shown below, the OPE coefficients  indeed contribute significantly to the final answer. Actually the twist operators in the product theory are not the truly local ones. More precisely, they always carry the disconnected multi-cycle structures in their definition such that the four-point correlation functions of twist operators are not single valued. 
    Nevertheless, there exist operator product expansions (OPE) of these twist operators as well. The price we pay is that the fusion rule for the OPE now is as follows
    \be \sigma_{g_{A}^{-1}} \sigma_{g_{B}} \to \sigma_{g_{B} g_{A}^{-1}}+...  \label{fusiontwist}\ee
   where the lowest-weight operator appearing in the expansion is not the identity operator, although $g_{A}$ and $g_{B}$ are conjugate to each other.  The ellipsis here denote the primary operators  with the same monodromy conditions as the twist operator $\sigma_{g_{B} g_{A}^{-1}}$ but with higher weights.
\end{itemize}

Now coming back to the evaluation about the reflected entropy $S_{R}(A:B)$, we can get directly from (\ref{nonabelian}) the weights of various twist operators 
\be h_{\sigma_{g_{A}}}=h_{\sigma_{g_{B}}}=nh_{m},\quad h_{\sigma_{g_{B}}\sigma_{g_{A}}^{-1} }=2h_{n}. \ee
This comes from the fact that  as we go through the replicated manifold with $mn$ sheets as shown in Figure \ref{fig:monodromy3}, we would find that $g_{A}$ and $g_{B}$ both contain $n$ $m$-cycles, and  $g_{B}g_{A}^{-1} $ contain 2 $n$-cycles. Next let us turn to the OPE coefficients in \eqref{fusiontwist}. We have 
\be \sigma_{g_{A}^{-1}} \sigma_{g_{B}} = C_{n,m} \sigma_{g_{B} g_{A}^{-1}}+...\quad , \quad \quad  \quad\mbox{with}\hs{2ex} C_{n,m}=(2m)^{-4h_{n}} \ee 
which can be proved by using the method presented in \cite{Lunin:2000yv,Dutta:2019gen}. 

\subsection{Reflected entropy: holographic 2D CFT} 


As is well known, in any 2D CFT with the Virasoro symmetries, a four-point correlation function of primary operators can be expanded by the Virasoro conformal blocks
\be 
\langle \cO_1(0)\cO_2(x)\cO_3(1)\cO(\infty)\rangle =\sum_{\cO_p}c_{12p}c_{34p}\cF(c,h_p,h_i,z)\bar{\cF}(c,\bar{h}_p,\bar{h}_i,\bar{z}) \label{blockexpcft}
\ee 
where $z=z_{12}z_{34}/z_{13}z_{24}$ is the standard cross ratio and we have expanded in $s$-channel, i.e. $z\to 0$ limit. Here $h_i,\bar{h}_i$ are the holomorphic and anti-holomorphic weights of the primary operator $\cO_i$, the summation is over all the primary operators $\cO_p$ appearing in the OPE, and $c_{ijp}$ denote the corresponding OPE coefficients. The Virasoro blocks $\cF$ capture all the contributions from the Verma module of $\cO_p$.  
In general, the Virasoro blocks do not have simple analytical expressions, and a way to manipulate them is to compute their forms in a series expansion by using the recursive relations first developed in \cite{Belavin:1984vu}. However, if we focus on the holographic CFT with a large central charge in the semi-classical limit, the Virasoro conformal block is expected to be exponentiated \cite{Belavin:1984vu,Hartman:2013mia}
\be 
\cF(c,h_p,h_i,z)\simeq \exp \left\{ -\frac{c}{6}f\left( \frac{h_p}{c},\frac{h_i}{c},z\right)\right\},
\ee 
where the function $f$ can be determined by the solution of certain well-defined monodromy problem. 

The holographic CFT is expected to own a sparse light spectrum, which would lead to the vacuum block dominance when calculating the four-point function of twist operators.
  In the following we would like to show the calculation of the reflected entropy $S_{R}(A:B)$ for two disjoint intervals sitting on the same time slice in both the $t$-channel and $s$-channel, and  in both the vacuum state and thermal state on the Lorentzian plane for later convenience. 

\paragraph{Vacuum state on the plane} 

Let us first consider the conformal block expansion of four-point correlation in the $t$-channel, i.e., $z\to 1$, which corresponds to the situation that the two disjoint intervals are quite close to each other compared with their own lengths. As shown in \eqref{blockexpcft}, 
\begin{align}
       & \langle \sigma_{{g}_A}(x_1) \sigma_{{g}^{-1}_A}(x_2) \sigma_{{g}_B}(x_3) \sigma_{{g}^{-1}_B} (x_4) \rangle _{\mbox{CFT}^{\otimes nm}(\mathbf{C})}\nn \\ 
      &=\frac{1}{x_{41}^{4h_{g_{A}}}x_{32}^{4h_{g_{A}}}} \sum_{\cO_p} C_{ABp}^{2} \mathcal{F}(mnc,h_{i},h_p,1-z) \overline{\mathcal{F}}(mn \bar{c},\bar{h}_{i},\bar{h}_p,1-\bar{z}) \nn\\
      &\approx \frac{1}{x_{41}^{ 4nh_{m}}x_{32}^{4nh_{m}}} C_{mn}^{2}   e^{-2 (2h_{n})  \ln{\frac{ 1+\sqrt{z} }{ 4(1-\sqrt{z}) }}}. \label{4pttwistt}
\end{align}
where $z$ is still the standard conformal invariant cross ratio, which is a real quantity in the symmetric setup. Now we are considering the product theory so the central charge are $mnc$ and $mn\bar{c}$. In the last step, we assume  the conformal block dominance from the single Virasoro block of the primary twist operator $\sigma_{g_{B} g_{A}^{-1}}$ in the fusion rule of OPE \eqref{fusiontwist}. In addition, we have used the fact that in the semi-classical limit defined by the conditions
\be mn c\to \infty, \qquad \text{while}\quad \frac{h_{i}}{mnc}, \frac{h_{p}}{mnc} \quad \text{fixed, } \ee
the Virasoro conformal block $\mathcal{F}$ would be exponentiated into the following form 
\be \log{\mathcal{F}(mnc,h_{i},h_p,1-z)} \sim -\frac{mnc}{6} f(\frac{h_{i}}{mnc},\frac{h_{p}}{mnc},1-z) \sim -h_{p}  \ln{\frac{ 1+\sqrt{z} }{ 4(1-\sqrt{z}) }}. \label{block1} \ee 
From the above expression we can find the reflected entropy for the vacuum state in 2D holographic CFTs at the leading order in the large $c$ limit
\begin{align}
    S_{R;vac}^{CFT}(A:B) 
    &=\frac{c}{3}\ln{\frac{ 1+\sqrt{z} }{ 1-\sqrt{z} }}. \label{division}
\end{align}
For convenience, we set the two disjoint intervals $A,B$ to be symmetric about the origin and sitting on a fixed time slice $t=0$
\be  A=(-b,-a),\quad B=(a,b),\quad z=(\frac{b-a}{b+a})^2, \label{interval} \ee  
then we have the final result: 
\be  S_{R;vac}^{CFT}(A:B) = \frac{c}{3}\ln\frac{b}{a}. \ee

\paragraph{Thermal state on the plane}

To represent the thermal state of 2D CFT at $t=0$ timeslice on a Lorentzian plane, we should construct an Euclidean path integral representation where the imaginary time direction should be compactified on an infinite long cylinder with circumference $\beta=T^{-1}$. Using the  map $z=e^{\frac{2\pi w}{\beta}}$, which takes the complex plane $(z,\bar{z})$ to the Euclidean cylinder $(w,\bar{w})$, we can 
get the resulting four point functions of twist operators on the thermal Euclidean cylinder  
 \bea 
     \lefteqn{ \langle \sigma_{{g}_A}(w_1) \sigma_{{g}^{-1}_A}(w_2) \sigma_{{g}_B}(w_3) \sigma_{{g}^{-1}_B} (w_4) \rangle _{\mbox{\small CFT}^{\otimes nm}}^{cylinder}} \notag \\
         &=&|z'(w_1)z'(w_2)z'w_3)z'(w_4)|^{2 h_{g_{A}}} \langle \sigma_{{g}_A}(z_1) \sigma_{{g}^{-1}_A}(z_2) \sigma_{{g}_B}(z_3) \sigma_{{g}^{-1}_B} (z_4) \rangle _{\mbox{\small CFT}^{\otimes nm}(\mathbf{C})}.
\label{cylinder1}
\eea 
Suppose we consider the symmetric configuration \eqref{interval} on the Euclidean cylinder at $\tau=0$ slice 
\be  A=(-b,-a), \quad B=(a,b), \label{Tconf} \ee 
which under the conformal mapping would lead to a configuration on the plane with a cross ratio
\be  z=\frac{z_{12}z_{34}}{z_{13}z_{24}}
       =\frac{\sinh{\frac{\pi z_{12}}{\beta}}\sinh{\frac{\pi z_{34}}{\beta}}}{\sinh{\frac{\pi z_{13}}{\beta}}\sinh{\frac{\pi z_{24}}{\beta}}} =\left( \frac{\sinh{\frac{\pi (b-a)}{\beta}}}{\sinh{\frac{\pi (b+a)}{\beta}}}\right)^2.
       \label{wcross}
\ee
Then  the reflected entropy on the thermal state $S_{R;thermal}^{CFT}(A:B)$ at the temperature $T=\beta^{-1}$ with the  configuration \eqref{Tconf} would be 
\begin{align}
      & S_{R;thermal}^{CFT}(A:B) =\frac{c}{3}\ln{\frac{ 1+\sqrt{z} }{ 1-\sqrt{z} }} =\frac{c}{3} \ln(\coth \frac{\pi a}{\beta} \tanh \frac{\pi b}{\beta}). 
\end{align}

\paragraph{Phase transition}

On the other hand, the $s$-channel conformal block expansion of the four point correlator \eqref{blockexpcft} comes from the OPE 
\be \sigma_{g_{A}} \sigma_{g_{A}^{-1}} \sim \textbf{1}+.... \ee
When taking the limit $x \to 0$ and the dominant Virasoro block is just the one related to the identity operator with dimension $h=\bar{h}=0$. Then the reflected entropies at the leading order of $c$ are simply
\be  S_{R;vac}^{CFT}(A:B)= S_{R;thermal}^{CFT}(A:B)=0.  \ee
As a result, there is a first-order phase transition when the cross ration $x$ goes from $0$ to $1$, which corresponds to the change of different dominant blocks in the conformal block expansion. This is similar to the situations happened in the mutual information $I(A:B)$ of two disconnected intervals.

\subsection{Entanglement wedge cross section}

It was proposed in \cite{Dutta:2019gen} that for a holographic CFT, the  reflected entropy $S_{R}(A:B)$ is captured holographically by the area of a  simple geometric object called the entanglement wedge cross-section (EWCS)
\be 
   S_{R}(A:B)= 2 \frac{Area(EWCS)}{4G}+ S_{R}(bulk)+..., \label{dictionary}
\ee
in which the first term is the semi-classic contribution in gravity and the other terms are from quantum corrections. 
In this subsection, we  give a brief review on  entanglement wedge and  entanglement wedge cross-section.     


The notion of entanglement wedge, which plays an essential role in the bulk reconstruction of AdS/CFT, was introduced in the literature when discussing the subregion/state  duality\cite{Czech:2012bh,Bousso:2012mh,Bousso:2012sj}, which aims to find the bulk region that is dual to the boundary reduced density matrix $\rho_{A B}$. In a generically time-dependent setting, the holographic entanglement entropy of a subregion $AB$ is given by the area of so-called HRT surface\cite{Hubeny:2007xt} $\g_{AB}$. The entanglement wedge
is the bulk domain of dependence of a co-dimension one surface, which  interpolates between $A B$ and the extremal surface $\gamma_{A B}$ anchored at $\partial (AB) $. 
Another closely related notion in the bulk reconstruction is the so-called bulk causal wedge, which is defined to be the intersection of the causal past and future of boundary domain of dependence $D[A B]$. In the case of pure AdS spacetime, these two wedges coincide with each other, while in more general situations, the entanglement wedge  contains the causal wedge.

\begin{figure}
    \centering
    \subfigure[]{
    \includegraphics[width=0.42\textwidth]{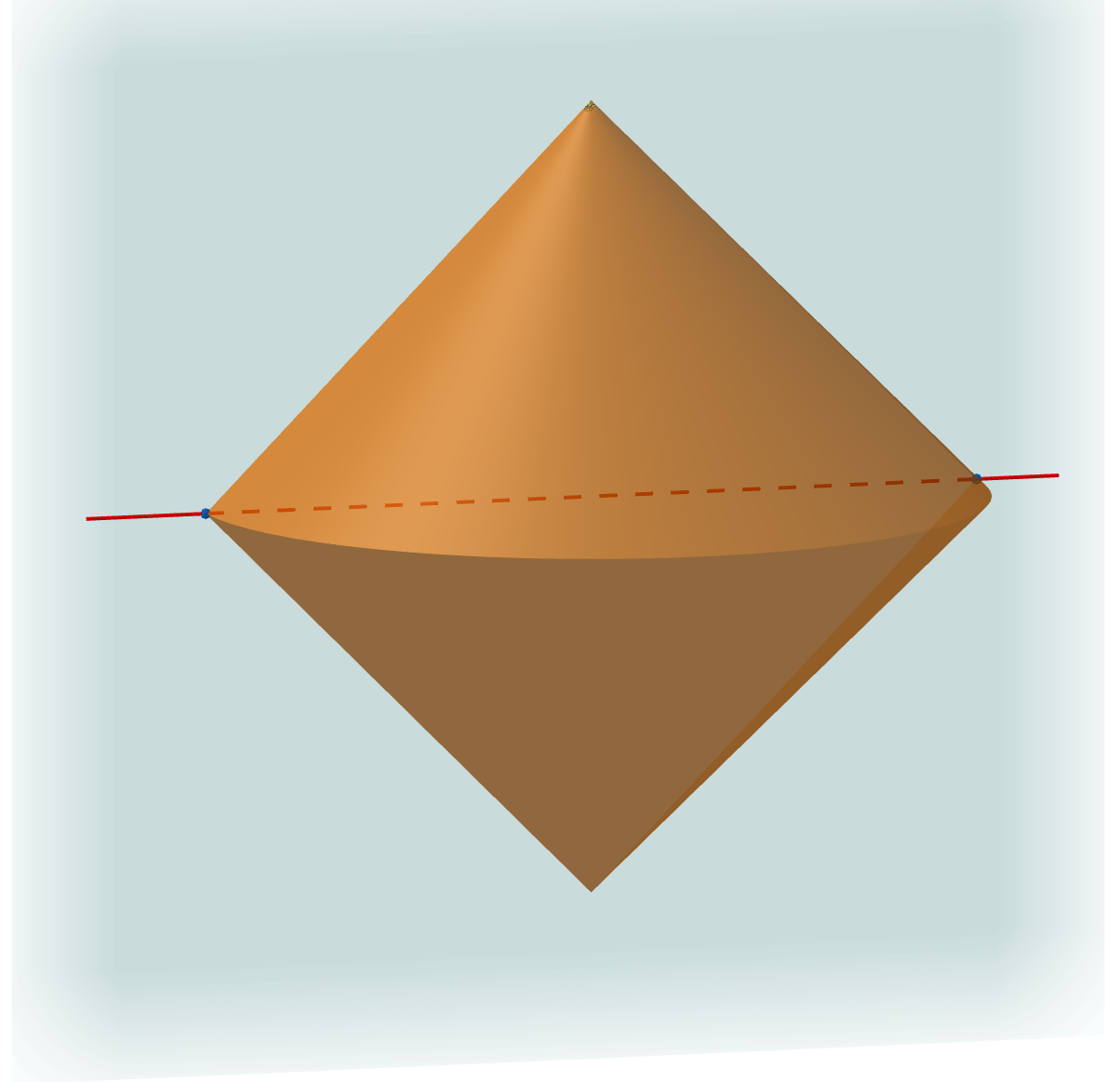}}
    \subfigure[]{
    \includegraphics[width=0.45\textwidth]{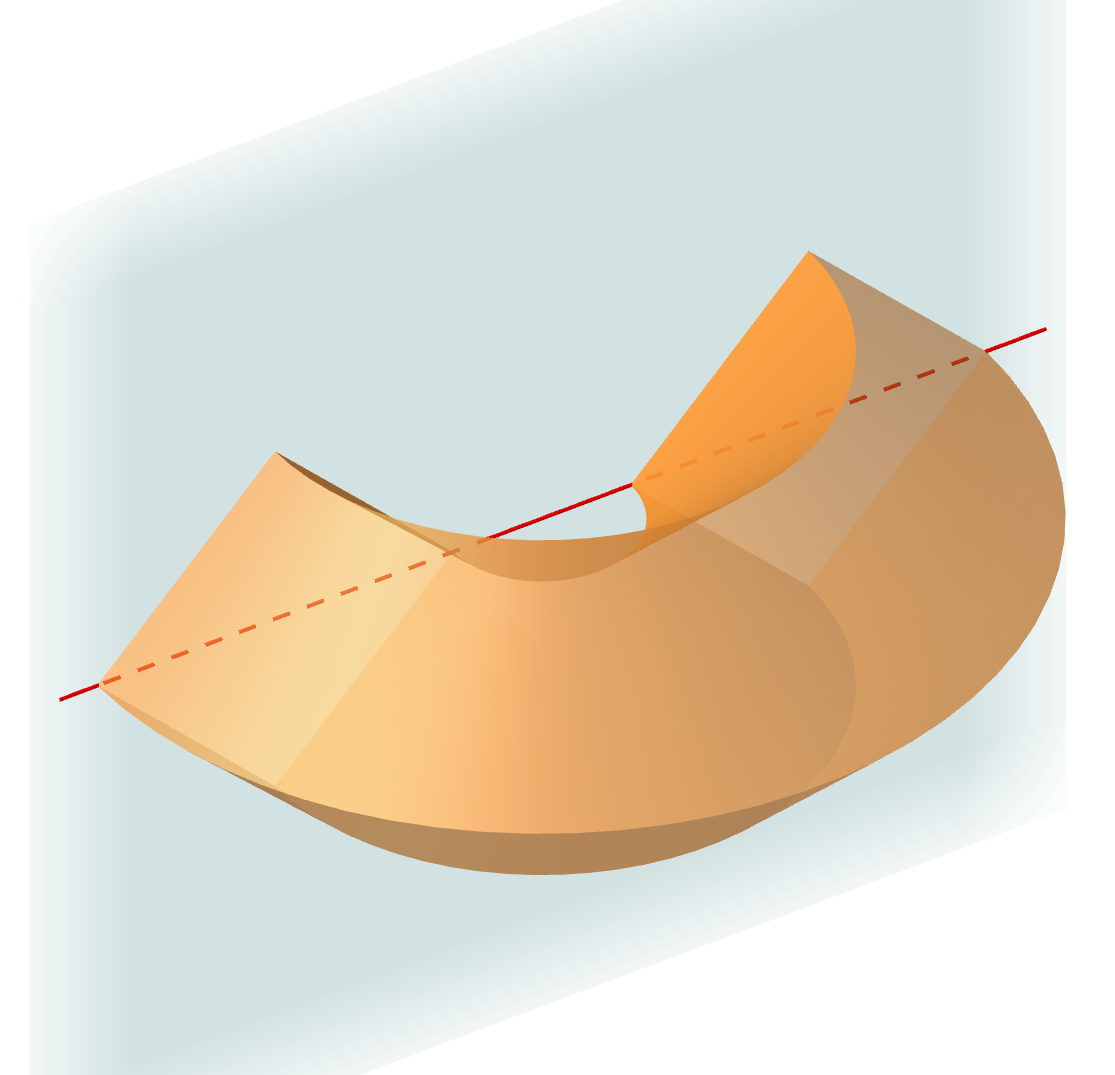}}
    \caption{Entanglement wedges related to single boundary interval and two boundary intervals (with large cross ratio) in Poincare coordinates of AdS$_3$. The blue plane denote the asymptotic boundary, the red line is the Cauchy surface on the boundary. The closed region bounded by yellow lightsheets and blue boundary is the entanglement wedge, whose intersection with the boundary is the domain of dependence of boundary subregion.}
    \label{fig:ewcft12}
    \end{figure}

The entanglement wedge cross-section (EWCS) is the minimal cross-section area of the entanglement wedge of two disconnected intervals. It is related to a bulk co-dimension 2 extremal surface whose endpoints lie on two separated null surfaces composing the boundary of entanglement wedge. Actually, this extremal surface can only exist on the surface $\Sigma_{AB}$ whose boundary is the union of the boundary intervals and its corresponding HRT surfaces $\partial \Sigma_{AB}= AB \cup \gamma_{AB}$, and its  endpoints must lie on two separated HRT surfaces. More generally, as a simple fact of differential geometry,  \textit{there is no extremal spacelike geodesics interpolating between two non-intersecting null surfaces, unless they have spacelike boundaries}.


\begin{figure}
    \centering
    \subfigure[]{
    \includegraphics[width=0.48\textwidth]{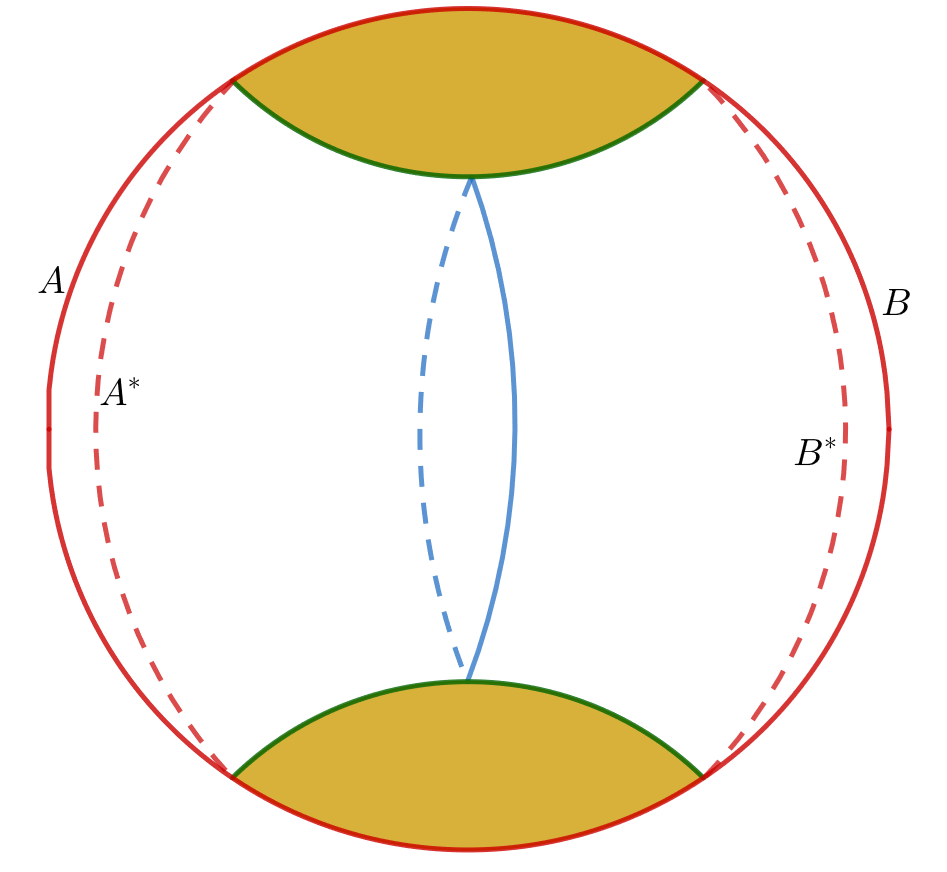}}
    \subfigure[]{
    \includegraphics[width=0.48\textwidth]{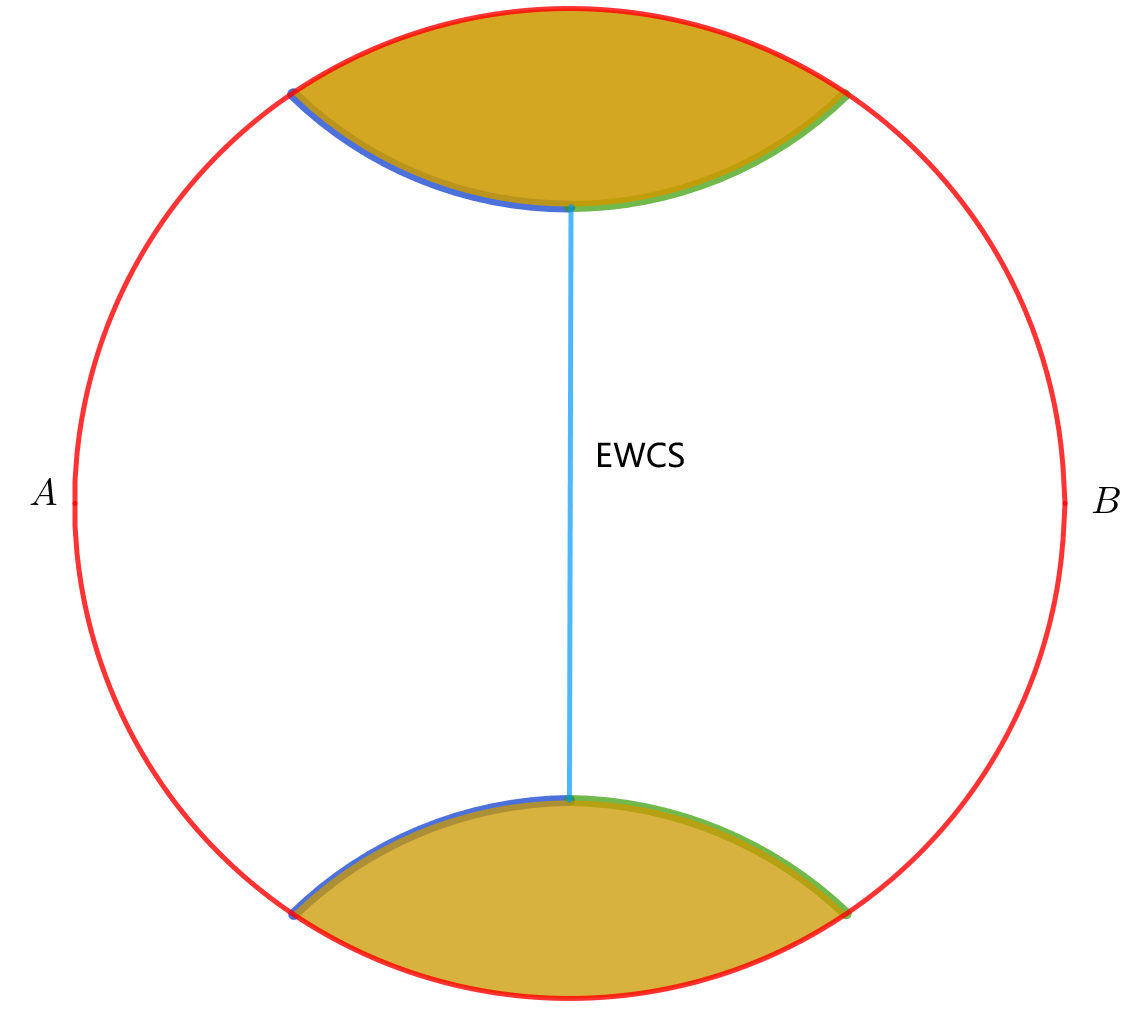}}
    \caption{Two dynamical constructions of  EWCS. The left figure (a) shows the cutting and gluing procedure. The field theory living on $AA^{*}$, $BB^{*}$ independently is the same as the original one. The Green lines denote $\gamma_{AB}$ which is the RT surface of $AB$. The right figure (b) shows that viewing the union of blue part of $\gamma_{AB}$ and $A$ as the new subregion (the union of green part of $\gamma_{AB}$ and $B$ is its complement), and EWCS is just the minimal RT surface.}
    \label{fig:twodynamical}
    \end{figure}


Although the kinematical definition of EWCS is clear, which is good enough to give the leading semi-classical  contribution in \eqref{dictionary}, there actually exist two dynamical constructions in the literature \cite{Takayanagi:2017knl,Nguyen:2017yqw,Dutta:2019gen}, which could lead to different first-order quantum corrections in \eqref{dictionary}. These two constructions both get intuition from the sub-region duality.  The first one \cite{Takayanagi:2017knl,Nguyen:2017yqw} take the RT surface $\gamma_{AB}$ as part of the new boundary and remove the geometry outside the entanglement wedge of
$AB$ so that the field theory defined on $ AB \cup \gamma_{AB} $ is in a pure state. Separating $\gamma_{AB}$ into two parts and combining $A$ or $B$ to get two regions, one may try to  minimize the entanglement between two regions by using the RT formula and then find that the minimal entanglement would give us the EWCS, as shown in Figure \ref{fig:twodynamical} (b). The second construction \cite{Dutta:2019gen} glues the entanglement wedge to its CPT conjugate using the Engelhardt-Wall procedure \cite{Engelhardt:2017aux,Engelhardt:2018kcs} and take the entangling surface  between the two boundary fields in this new two-sided black hole  as twice the EWCS, as shown in  Figure \ref{fig:twodynamical} (a).




\section{Reflected entropy in 2D WCFT \label{fieldside}}

 In the following sections we would like to consider the reflected entropy in the framework of AdS$_3$/WCFT\cite{Compere:2013bya}.  As it has been shown in \cite{Song:2016gtd,Apolo:2020bld}, the holographic entanglement entropy in this correspondence presents some novel features, it would be interesting to investigate the holographic description of the reflected entropy. As the first step, we study the computation of the reflected entropy in 2D warped conformal field theories in this section. To compute the reflected entropy, we have to study carefully the properties of the relevant twist operators in the orbifold theory of WCFT. 
 \subsection{Warped conformal field theory}
 
2D WCFT is a non-Lorentzian invariant field theory, which possess an infinite dimensional symmetry
named the warped conformal symmetry. The symmetry is generated by a Virasoro algebra plus a $U(1)$ Kac-Moody algebra, with the global part being $SL(2,R) \times U(1)$. This warped conformal symmetry is constraining and powerful, and allows us to obtain the Cardy-like  formula \cite{Detournay:2012pc}, one-loop partition functions \cite{Castro:2015uaa} and  to do the modular bootstrap\cite{Apolo:2018eky}.
 Let us first consider a 2D WCFT on the reference plane with coordinates $(x,y)$. In position space a general warped conformal transformation can be written as 
\be \label{warped-trans}
x'=f(x),\quad y'=y+g(x) \ee
where $f(x)$ and $g(x)$ are two arbitrary functions. Let $T(x)$ and $P(x)$ be the Noether currents associated with the translations along $x$ and $y$ directions respectively. Under the warped conformal transformations \eqref{warped-trans}, their transformation properties are given by
\be 
\begin{split}
    T'(x')&=\Big(\frac{\partial x}{\partial x'}\Big)^2\Big( T(x)-\frac{c}{12}\{f,x\} \Big)+ \frac{\partial x}{\partial x'}\frac{\partial y}{\partial x'}P(x)-\frac{k}{4} \Big(\frac{\partial y}{\partial x'}\Big)^2, \\
    P'(x')&= \frac{\partial x}{\partial x'}\Big( P(x)+\frac{k}{2}\frac{\partial y'}{\partial x} \Big),
\end{split}
\label{wcfttransf1}
\ee 
where $ \{f,x\}$ denotes the standard Schwarzian derivative, $c$ is the central charge and $k$ is the $U(1)$ Kac level. We can construct infinitely many conserved charges out of these two currents
\begin{equation}
    \begin{aligned}
    L_n=-\frac{i}{2\pi}\int dxx^{n+1}T(x),\quad P_n=-\frac{1}{2\pi}\int dxx^nP(x).
    \end{aligned}
\end{equation}
The commutation relations for these charges form a warped conformal algebra consisting of a Virasoro algebra and a Kac-Moody algebra,
\begin{equation}
    \begin{aligned}
    [L_{n},L_{m}]&= (n-m)L_{n+m}+\frac{c}{12}n(n^2-1)\delta_{n,-m}, \\
    [L_n,P_m]&=-mP_{n+m}, \\
    [P_n,P_m]&=k\frac{n}{2}\delta_{n,-m}.
\end{aligned} 
\end{equation}

When describing 2D WCFT, it is important to account for spectral flow transformations that induce a redefinition of charges $L_n$ and $P_n$ while leaving the Virasoro-Kac-Moody algebra unchanged. On the reference plane $(x,y)$ of WCFT the vacuum expectation values of conserved currents are zero $ \langle T(x) \rangle=\langle P(x) \rangle=0$, which means that this vacuum state is invariant under the global $SL(2,R) \times U(1)$ transformation. Moreover the vacuum state on the reference plane satisfies
\be L_{n}|0\rangle=P_{n+1} |0\rangle =0,\quad n \ge -1. \ee
We can also define WCFT on a canonical plane with coordinates $(x',y')$ by introducing a spectral flow parameter $\mu$ together with a specific warped conformal transformation 
\be x'= x, \quad y'=y+\mu \log{(x)}. \label{reftocan1} \ee
Then the spectral flowed charges $L'_{n}$ and $P'_{n}$ satisfy following relations,
\be L'_{n}=L_{n}-i\mu P_{n}-\frac{\mu^2 k}{4} \delta_{n}, \quad P'_{n}=P_{n}-\frac{i \mu k}{2}\delta_{n} \label{tr2} \ee 
which lead to the non-vanishing vacuum expectation values of zero modes $L'_0$ and $P'_0$ on this canonical plane
\be \langle L'_{0} \rangle=-\frac{\mu^2 k}{4}, \quad  \langle P'_{0} \rangle=-\frac{i \mu k}{2}. \ee 
Note that if the spectral flow parameter $\mu$ is real, then the charge $\langle P'_{0} \rangle$ is imaginary, which is a special feature of WCFT with holographic dual. 

To match with the holographic calculations in asymptotic AdS$_{3}$ spacetime, we need to define WCFT having specific vacuum charges on the canonical cylinder with the coordinates $ (\hat{x}',\hat{y}') $. This can be accomplished by using an exponential map in the  direction $x'$, while keeping the direction $y'$ unchanged
\be x'=e^{i \hat{x}'}, \quad \hat{y}'=y'.  \ee
The vacuum charges on this canonical cylinder can be obtained by adding a term $-\frac{c}{24}$ to the Virasoro zero mode on the canonical plane, like what happens in 2D CFT. To be complete, we can also define the reference cylinder $(\hat{x},\hat{y})$ by the same exponential map related to the reference plane. These two cylinders are related by a spectral flow transformation 
\be  \hat{x}=\hat{x}', \quad \hat{y}=\hat{y}'-i \mu \hat{x}'   \ee  
where the parameter $\mu$ is the same one as in \eqref{reftocan1}. Because of boundary conditions, we start from the canonical cylinder with the identification $ (\hat{x},\hat{y}') \sim (\hat{x}+2\pi ,\hat{y}') $, which leads to a corresponding identification on the reference plane
\be (x,y) \sim (x e^{2\pi i},y-2\pi i \mu). \label{refplaneb} \ee
Note that \eqref{refplaneb} is consistent with the modular flow transformation \eqref{reftocan1}.

Different setups of the WCFT play their own specific roles in exploring various properties of WCFT and its holographic dual. We summarize here for later convenience, 

\begin{itemize}
    \item Reference plane:  \be (x,y) \sim (x e^{2\pi i},y-2\pi i \mu),\quad \langle L^{vac}_{0} \rangle=0,\quad \langle P^{vac}_{0} \rangle=0.  \label{refp0} \ee 
    
    \item Canonical plane:  \be (x',y') \sim (x' e^{2\pi i},y'),\quad \langle L^{vac}_{0} \rangle=-\frac{\mu^2 k}{4},\quad \langle P^{vac}_{0} \rangle=-\frac{i\mu k}{2}. \label{can11}\ee
    
     \item Reference cylinder:  \be (\hat{x},\hat{y}) \sim (\hat{x}+2\pi ,\hat{y}-2\pi i \mu),\quad \langle L^{vac}_{0} \rangle=-\frac{c}{24},\quad \langle P^{vac}_{0} \rangle= 0.\ee 
     
     \item Canonical cylinder:  \be (\hat{x}',\hat{y}') \sim (\hat{x}'+2\pi ,\hat{y}'),\quad \langle L^{vac}_{0} \rangle=-\frac{c}{24}-\frac{\mu^2 k}{4},\quad \langle P^{vac}_{0} \rangle= -\frac{i\mu k}{2}. \label{can cyl} \ee
     
     \item The coordinate transformations among these setups are
     \bea x=x', && y=y'- \mu \log(x'), \nn  \\
      \hat{x}=\hat{x}',&& \hat{y}=\hat{y}'- i \mu \hat{x}', \nn  \\
      x=e^{i \hat{x}}, && y=\hat{y}. \eea
\end{itemize}
Note that as the underlying geometry of WCFT is not the usual Riemann surfaces,  neither the plane nor the cylinder in which a WCFT is defined is the complex plane or Euclidean cylinder in the usual sense.

Another remarkable point is that there exist spectral-flow invariant Virasoro generators
\be  L^{inv}_{n}=L_{n}-\frac{1}{k} \big(\sum_{m\le -1} P_{m}P_{n-m}+\sum_{m\ge 0} P_{n-m}P_{m}  \big),  \label{flowinv} \ee
which commute with all the Kac-Moody generators $P_n$, and form a Virasoro algebra with central charge $c-1$. A primary state is defined to be in the highest weight representation, and it is characterized by the weight $h$ and the charge $q$, which are eigenvalues of $L_{0}$ and $P_{0}$ respectively. Then the primary state $|h,q \rangle$ under $L_{n}$ and $P_{n}$ is also a primary state under $L_{n}^{inv}$ and $P_{n}$, but with the conformal weight being shifted by a $q$-dependent quantity
\be  h^{inv}=h-\frac{q^2}{k}. \label{wcftinv} \ee 
Since the spectral flow invariant Virasoro $L_{n}^{inv}$ commute with $P_{n}$, it is more convenient to label the state basis, i.e. the primary states and their warped conformal descendants, by $(h^{inv},q)$ when considering the conformal block expansion.


Due to the vanishing of the vacuum charges on the reference plane, we can completely determine the two- and three-point functions on the vacuum state of WCFT by using the global $SL(2,R)\times U(1)$ symmetries. Moreover, the four-point function is determined up to a function of the cross-ratio $ \mathrm{x}=\frac{x_{12}x_{34}}{x_{13}x_{24}}$ with $x_{ij}\equiv x_{i}-x_{j}$,
\be \langle \phi_{1} \phi_{2} \phi_{3}\phi_4 \rangle=\big(e^{i\sum_{j}q_{j}y_{j}} \delta({\sum_{i}q_{i}}) \big) \big( \frac{x_{24}}{x_{14}}\big)^{h_{12}}\Big( \frac{x_{14}}{x_{13}}\Big)^{h_{34}} \frac{\mathcal{G}(\mathrm{x})}{x_{12}^{h_{1}+h_{2}} x_{34}^{h_{3}+h_{4}} }, \label{wcft4p} \ee
where  $\phi_i \equiv \phi_{i}(x_{i},y_{i})$ is the primary operator with conformal weight $h_{i}$ and $U(1)$ charge $q_{i}$ respectively,  and $\mathcal{G}(\mathrm{x})$ is an undetermined function of the cross ratio $\mathrm{x}$. 
The function $\mathcal{G}(\mathrm{x})$ can be decomposed into the warped conformal blocks $A^{h,q}(\mathrm{x})$
\be \mathcal{G}(\mathrm{x})=\sum_{\mathcal{O}} C_{12\mathcal{O}} C^{\mathcal{O}}_{34} A^{h,q}(\mathrm{x}),  \ee
where the sum runs over the primary states with weight $h$ and charge $q=q_{3}+q_{4}=-q_{1}-q_{2}$, and $C_{12\mathcal{O}},C^{\mathcal{O}}_{34}$ are OPE coefficients. Actually the full warped conformal block is just the Virasoro-Kac-Moody block whose expression has been found in chiral CFTs with an internal $U(1)$ symmetry in \cite{Fitzpatrick:2015zha}.  More precisely, the warped conformal block can be factorized as  
\begin{align}
    A^{h,q}(\mathrm{x})= \mathcal{V}_{p}(k,q_{i},\mathrm{x}) \mathcal{F}(c-1,h_{i}^{inv},h^{inv},\mathrm{x}) = \mathrm{x}^{\frac{2q_{1}^2}{k}}(1-\mathrm{x})^{\frac{2q_{1}q_{2}}{k}}\mathcal{F}(c-1,h_{i}^{inv},h^{inv},\mathrm{x})
    \label{wcft 4point}
\end{align}
where $\mathcal{V}_{p}(k,q_{i},\mathrm{x})$ is the Kac-Moody block, $\mathcal{F}(c-1,h_{i}^{inv},h^{inv},\mathrm{x})$ is the standard Virasoro conformal block with central charge $c-1$. 
 
Under the warped conformal transformation \eqref{warped-trans}, the $n$-point functions of primary fields transform as
\begin{equation}\label{npt-trans}
    \langle\prod_{i=1}^n\phi'_i(x'_i,y'_i)\rangle=\prod_{i=1}^n\Big(\frac{\partial x'_{i}}{\partial x_i}\Big)^{-h_i} \langle\prod_{i=1}^n\phi_i(x_i,y_i)\rangle.
\end{equation}
Note that it only depends on the diffeomorphism of $x$.

\subsection{Twist operators in orbifold WCFT}

To compute the reflected entropy, it is essential to understand the properties of twist operators in non-Abelian orbifold WCFT. Let us first consider the  Abelian orbifold, which appears in the study of entanglement entropy. 

\subsubsection{Twist operator in Abelian orbifold WCFT }

In computing the single-interval entanglement entropy, one way is to compute the two-point function of twist operators. In this case, the resulting orbifold is an Abelian orbifold. 
The conformal weight $h_n$ and the $U(1)$ charge $q_n$ of the twist field $\sigma_{n}$ for WCFT were determined in \cite{Castro:2015csg,Song:2017czq} by using a generalization of Rindler method developed in \cite{Casini:2011kv}. The results are as follows
\be h_{n}=n(\frac{c}{24}+\frac{L_{0}^{vac}}{n^2}+\frac{i P_{0}^{vac} \alpha}{2n \pi}-\frac{\alpha^2 k}{16 \pi^2}), \quad q_{n}=n(\frac{P_{0}^{vac}}{n}+i\frac{ k \alpha}{4\pi}),   \label{h-q-wcft} \ee
where $L_{0}^{vac}$ and $P_{0}^{vac}$ denote the vacuum expectation values of $L_{0}$ and $P_{0}$ on the cylinder respectively. The variable $\alpha$ was introduced in \cite{Song:2016gtd} as a free parameter when using the generalized Rindler method, and its  value was later determined by matching the single interval boundary entanglement entropy  with the holographic ones in BTZ black holes. In the $n \to 1$ limit,  the Abelian orbifold WCFT reduces to the original theory, which require that $h_{1}=q_{1}=0$. This fact implies the values of the vacuum charges
\be L_{0}^{vac}=-\frac{c}{24}-\frac{\alpha^2 k}{16 \pi^2}, \quad P_{0}^{vac}=-\frac{i\alpha k}{4 \pi}  \label{charge-vac},  \ee 
which is precisely the spectral flowed charges defined on the canonical cylinder \eqref{can cyl} if we make an identification between the parameter $\alpha$ and the spectral flow parameter $\mu$ \eqref{reftocan1} of the WCFT  
\be \mu=\frac{\alpha}{2 \pi}  \label{mu-alpha}. \ee
Substituting \eqref{charge-vac} and \eqref{mu-alpha} into \eqref{h-q-wcft}, we find 
\be h_{n}=\frac{n-1}{n} [(n+1)\frac{c}{24}-(n-1)\frac{\mu^2 k}{24}],\quad q_{n}=(n-1)\frac{i \mu k}{2}, \label{chargeRindler}\ee
which are proportional to $n-1$ as expected. For later convenience, we mention two things here:
\begin{itemize}
    \item In the conformal block expansion of the 4-point function \eqref{wcft 4point}, we use $h^{inv}$ as the convenient parameter because of its invariance under the spectral flow. In the case at hand, there is 
    \be h_{n}^{inv}=h_{n}^2-\frac{q_{n}^{2}}{k}=\frac{c}{24}\frac{n^2-1}{n}, \ee
    which is exactly the holomorphic weight of the twist operator in 2D orbifold CFT. This relation will help us explicitly see the entropy relations between CFT and WCFT. 

    \item  To reproduce the values of $L_{0}^{vac}$ and $P_{0}^{vac}$ found in holographic entanglement entropy calculation of WCFT (where $k$ is assumed to be negative), the parameter $\mu$ must take the following value 
    \be \mu=\sqrt{\frac{-c k}{6}} \quad \to \quad   L_{0}^{vac}=0,\quad  P_{0}^{vac}=\frac{i}{2} \sqrt{\frac{-c}{6k}}. \label{match12} \ee 
\end{itemize}

By using \eqref{h-q-wcft},  we can get the entanglement entropy of single interval $\cal A$ with endpoints $\partial\mathcal{A}=\{(x_1,y_1),(x_2,y_2)\}$ on the reference plane
\be S_{\mathcal{A}}=-i P_{0}^{vac}y_{21}-2(i\mu P_{0}^{vac}+2L_{0}^{vac})\log{\left( \frac{x_{21}}{\epsilon} \right)},\ee
where the parameter $\epsilon$ is a UV regulator in the $x$ direction.
In contrast, the entanglement entropy of the same interval $\cal A$ on the canonical plane is
\be S_{ \mathcal{A} }=-i P_{0}^{vac}\left( y'_{21}-\mu \log{\frac{x_2}{x_1}} \right) -2 \left( i\mu P_{0}^{vac}+2L_{0}^{vac} \right) \log{ \left( \frac{x_{21}}{\epsilon}\right) }.  \label{vacuument}\ee

 To compute the entanglement entropy of single interval at a finite temperature, we consider WCFT on the thermal cylinder with the general coordinate identification $(w^{+},w^{-})\sim (w^{+}+i \beta,w^{-}-i \bar{\beta} )$, which can be obtained via a special warped conformal transformation from the reference plane
\be x=e^{\frac{2\pi}{\beta}w^{+}},\quad y=w^{-}+\frac{\bar{\beta}-2\pi \mu}{\beta}w^{+}. \label{wcftT} \ee
Using the transformation property of correlation functions \eqref{npt-trans}, we obtain 
\be S_{\mathcal{A}}^{\beta,\bar{\beta}}=-i P_{0}^{vac}\left(\bar{l}+\frac{\bar{\beta }-2 \pi \mu}{\beta}l\right)-2(i\mu P_{0}^{vac}+2L_{0}^{vac})\log{ \big(\frac{\beta}{\pi \epsilon}\sinh{\frac{\pi l}{\beta}} \big)},  \label{thermalent} \ee
where $l=|w^{+}_{21}|$ and $\bar{l}=|w^{-}_{21}|$. Note that when taking the zero-temperature limit of  \eqref{thermalent}, it does not simply reduce to the vacuum one \eqref{vacuument} due to the presence of the spectral flow parameter $\mu$.  

\subsubsection{Twist operators in non-Abelian orbifold WCFT }

As reviewed in section 2, to compute the reflected entropy, one may consider the correlation function of twist operators in a non-Abelian orbifold. Different from the entanglement entropy case, the twist operators here are more subtle. In this subsection, our goal is to give a detailed analysis on the field identifications induced by different twist operators in the orbifold WCFT, and find a way from these specific monodromy conditions to calculate the quantum numbers of the twist operators that should be consistent with the charge conservation. We would give several non-trivial examples to show the effectiveness of our method. 

In 2D CFT, the holomorphic and anti-holomorphic weight of local primary twist and anti-twist operators $\sigma_{p}$ and $\Tilde{\sigma}_{p}$ locating at the endpoints of the subregions can be determined by evaluating the expectation values of the conserved current $T(X)$ on the replicated Riemannian manifold  $\Sigma_{r}$ 
\be \langle T(z^{j},\Bar{z}^{j}) \rangle_{\Sigma_{r}}=\frac{\langle T(z,\Bar{z}) \sigma_{r}(z_{1},\Bar{z}_{1}) \Tilde{\sigma}_{r}(z_{2},\Bar{z}_{2})\rangle_{\mbox{CFT}(\cal C) } }{\langle \sigma_{r}(z_{1},\Bar{z}_{1}) \Tilde{\sigma}_{r}(z_{2},\Bar{z}_{2}) \rangle_{\mbox{CFT}(\cal C)}} \label{cftstress} \ee 
where $T(z^{j})$ represents the stress tensor on the $j$-th replica sheet. The left-hand side can be evaluated by the
transformation rule of the stress tensor $T(z)$ under a uniformization map from replicated manifold $\Sigma_{r}$ to the complex plane $\mathcal{C}$. The other side can be evaluated by using the Ward identity and the two-point correlator of the twist operators. A 2D CFT is defined on Riemann surfaces with a complex structure. Thus the anti-holomorphic coordinate $\Bar{z}$ must transforms in  accord with the holomorphic one $z$. More explicitly when $z \to z e^{2 \pi i}$, there must be $\Bar{z} \to \Bar{z} e^{-2 \pi i}$. Due to this relation, the uniformization map about anti-holomorphic coordinate is similar to the ones of holomorphic coordinate, and so is the anti-holomorphic weight.

  
 


\begin{figure}
    \centering
    \includegraphics[width=.8\textwidth]{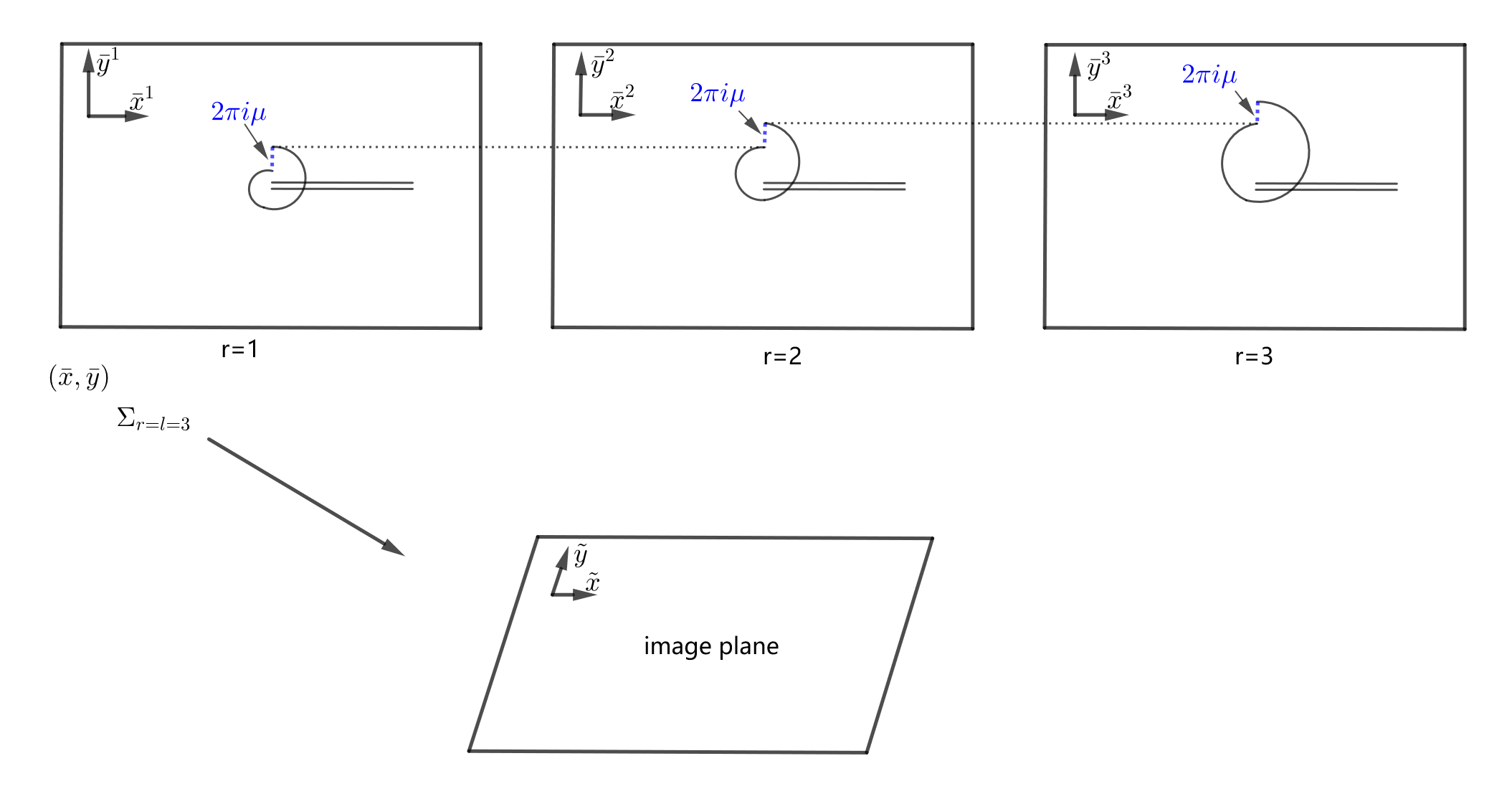}
    \caption{Three physical sheets in coordinates $(\bar{x}^{j},\bar{y}^{j})$ with $j=1,2,3$ constitute the replicated manifold $\Sigma_{m=3}$. When the parameter $p$ in \eqref{uniformap} is zero, each physical sheet is a reference plane with identification $(\bar{x}^{j},\bar{y}^{j}) \sim (\bar{x}^{j} e^{2 \pi i}, \bar{y}^{j}-2\pi i \mu )$. The dotted blue lines show the shifts in $y$ direction on each replica sheet, and the dotted black lines help us understand their relations. Under a uniformization map \eqref{uniformap}, the image plane with coordinates $(\Tilde{x},\Tilde{y})$ can be chosen as a reference plane, i.e., $\Tilde{y}=y$ where $y$ is defined in \eqref{refp0}, or a canonical plane, i.e., $\Tilde{y}=y'$ where $y'$ is defined in \eqref{can11}. Here we choose the image plane as a reference plane for simplicity.}
    \label{fig:uniformap}
\end{figure}

In 2D WCFT, the local primary twist operator is a charged field under the $U(1)$ Kac-Moody current. Thus, the Ward identities related to the holomorphic energy momentum tensor $T(X)$ and the $U(1)$ current $P(X)$ would help us to find the  conformal dimensions and charges of various twist fields. 
  Analogues to the case of 2D CFT, we determine the dimensions and charges of the twist operators in WCFT by computing the expectation value of the stress tensor $T(\bar{x}^i)$ and the current $P(\bar{x}^i)$ on the replicated manifold $\Sigma_r$ 
\begin{align}
  &\langle T(\bar{x}^i) \rangle_{\Sigma_{r}}=\frac{\langle T(\bar{x}) \s_{g_{r}}(\bar{x}_1,\bar{y}_1) \Tilde{\sigma}_{g_{r}}(\bar{x}_2,\bar{y}_2)\rangle_{\mathcal{C}_p } }{\langle \s_{g_{r}}(\bar{x}_1,\bar{y}_1) \Tilde{\sigma}_{g_{r}}(\bar{x}_2,\bar{y}_2) \rangle_{\mathcal{C}_p}}, \nn \\
  &\langle P(\bar{x}^i) \rangle_{\Sigma_{r}}=\frac{\langle P(\bar{x}) \s_{g_{r}}(\bar{x}_1,\bar{y}_1) \Tilde{\sigma}_{g_{r}}(\bar{x}_2,\bar{y}_2)\rangle_{\mathcal{C}_p } }{\langle \s_{g_{r}}(\bar{x}_1,\bar{y}_1) \Tilde{\sigma}_{g_{r}}(\bar{x}_2,\bar{y}_2) \rangle_{\mathcal{C}_p}} \label{wcft-stress-charge}
\end{align}
where $r  $ denotes the replica number of the twist operators $\s_{g_{r}}$ and $g_r$ is the group element in the definition of twist operator $\s_{g_{r}}$, which in the Abelian orbifold case equals to $(12...r)$ for the twist operator (the anti-twist operator is related to the group element $(r (r-1)...1)$). 
However,  unlike  2D CFT, the background geometry of 2D WCFT is a type of Newton-Cartan geometry \cite{Hofman:2014loa,Castro:2015csg} with a scaling structure $J^{a}_{b}$. 
The two axes $x$ and $y$ are independent with each other. The two subgroups of global group $SL(2,R)\times U(1)$ of a 2D WCFT act on $x$ and $y$ respectively. 
 For the coordinate $x$, it plays a similar role in 2D WCFT as the holomorphic coordinate $z$ in complex plane of 2D CFT.  While for the other axis $y$,  it behaves in a special manner under the spectral flow. Actually, it needs special attention to find the transformation of the $y$ coordinate in defining  various twist operators.
Moreover, there is another subtlety in computing the dimensions and charges of twist operators in WCFT by using the uniformization map method. Due to the distinct property of Newton-Cartan geometry on which WCFT are defined, there is a whole family of planes $\cC_p$ parametrized by the spectral flowed parameter $p$ with different vacuum expectation values of conserved charges. The coordinates $(\bar{x},\bar{y})$ on  $\cC_p$ are related to the coordinates $(x,y)$ on the reference plane by 
\be \bar{x}=x, \quad \bar{y} =y- p \, x.  \label{trnas1}\ee
More concretely, there are three kinds of geometries appearing in our calculation, see Figure \ref{fig:uniformap}: 1). The original physical plane $\cC_p$ with coordinates $(\bar{x},\bar{y})$, whose properties are determined by parameter $p$;  2). The replicated geometry $\Sigma_{r}$ with coordinates $(\bar{x}^{i},\bar{y}^{i}),\,i=1...r$, which is composed of $r$ original physical planes $\cC_{p}$;     3). The image plane $\cC_{img}$ with coordinates $(\Tilde{x},\Tilde{y})$, which could be chosen as a canonical plane, a reference plane or other more general spectral flowed ones under the uniformization map \eqref{uniformap}.

In evaluating the left-hand side of \eqref{cftstress}, the replicated manifold $\Sigma_{r}$ and the image plane $\mathcal{C}_{img}$  are relevant. While in order to evaluate the right-hand side of \eqref{cftstress}, we need to use the physical plane $\cC_p$ where the original WCFT lives, because we have
\be \Sigma^{r}/\mb{Z}^{r}=\text{Original physical plane} \;\; \cC_p \ee
Thus the vacuum charges related to the operators $T$ and $P$ are distinct on the evaluation of the two sides of \eqref{cftstress}, which would be shown explicitly in the following calculations.    

  Let us evaluate the right-hand side of \eqref{wcft-stress-charge}. The two-point function on the physical plane $\cC_p$ has the following form
\be  \langle \phi_{1} \phi_{2} \rangle_{ \mathcal{C}_{p}}=\big(e^{i\sum_{j}q_{j}( \bar{y}_{j}+p \bar{x}_{j}) }  \delta({\sum_{j}q_{j}}) \big) \frac{\delta (h_{\phi_{1}}- h_{\phi_{2}})}{(\bar{x}_{12})^{2h_{\phi_{1} }}} , \label{2pointflow}  \ee 
and the vacuum expectation values of charges  are given by
\be  \langle T(\bar{x}) \rangle_{\mathcal{C}_{p}}=-\frac{k p^2}{4}, \quad \langle P(\bar{x}) \rangle_{\mathcal{C}_{p}}=-\frac{k p }{2}. \ee
Note that the above result can be computed by \eqref{trnas1}, \eqref{wcfttransf1} and \eqref{tr2}. The Ward identities for the energy momentum tensor $T(\bar{x})$ and the $U(1)$ current $P(\bar{x})$ on the physical plane $\mathcal{C}_{p}$ are \cite{Belavin:1984vu}
\begin{align}\label{wcft-ward}
   &  \langle T(\bar{x}) \sigma_{r}(\bar{x}_{1},\bar{y}_{1}) \Tilde{\sigma}_{r}(\bar{x}_{2},\bar{y}_{2}) \rangle_{\mathcal{C}_{p} }\nn \\
  &\hs{5ex}=  \sum_{i=1}^{2} \left( \frac{\Delta_{r}/r}{(\bar{x}-\bar{x}_{i})^{2}}+ \frac{1/r}{ \bar{x}-\bar{x}_{i} } \frac{\partial}{\partial \bar{x}_{i}} +\langle T(\bar{x}) \rangle_{\mathcal{C}_{p}} \right)  \langle  \sigma_{r}(\bar{x}_{1},\bar{y}_{1}) \Tilde{\sigma}_{r}(\bar{x}_{2},\bar{y}_{2}) \rangle_{ \mathcal{C}_{p}}, \nn \\
   &  \langle P(\bar{x}) \sigma_{r}(\bar{x}_{1},\bar{y}_{1}) \Tilde{\sigma}_{r}(\bar{x}_{2},\bar{y}_{2}) \rangle_{ \mathcal{C}_{p}} \nn \\
  &\hs{5ex}=  \sum_{i=1}^{2} \left(   \frac{1/r}{ \bar{x}-\bar{x}_{i} } \frac{\partial}{\partial \bar{y}_{i}} +\langle P(\bar{x}) \rangle_{ \mathcal{C}_{p}} \right)  \langle  \sigma_{r}(\bar{x}_{1},\bar{y}_{1}) \Tilde{\sigma}_{r}(\bar{x}_{2},\bar{y}_{2}) \rangle_{ \mathcal{C}_{p}}. 
\end{align}
 We place the twist operator $\s_r$ and the anti-twist operator $\Tilde{\s}_r$ at points $(0,0)$ and $(l_0,0)$ separately.
Substituting the two-point function \eqref{2pointflow} and the Ward identities \eqref{wcft-ward} into \eqref{wcft-stress-charge}, we find
\begin{align}
    \langle P(\bar{x}^{i}) \rangle_{\Sigma_{r}}&= \frac{l_0}{\bar{x}(\bar{x}-l_0)} \frac{-i q_{r}}{r}-\frac{kp}{2},
    \label{pcharge1}\\
    \langle T(\bar{x}^{i}) \rangle_{\Sigma_{r}}&= \frac{l_{0}^{2} }{\bar{x}^2(\bar{x}-l_{0})^2} \frac{h_{r}}{r} +p\frac{l_{0}}{\bar{x}(\bar{x}-l_{0})} \frac{ -i q_{r}}{r} -\frac{k p^2}{4} \label{tcharge1}
\end{align}
where $q_r$, $h_r$ are undetermined quantum numbers of the twist operator $\s_{r}$.

Now we turn to the computation of left-hand side of \eqref{wcft-stress-charge}. The  uniformization map has the form
\be  \Tilde{x}=\left( \frac{\bar{x} }{\bar{x}-l_0} \right)^{\frac{1}{r}}, \quad   \Tilde{y} -\mathrm{t} \mu \log{\Tilde{x}}+ \kappa \mu \log{\Tilde{x}} =\bar{y}+p \bar{x}  \label{uniformap}\ee 
where $\mu$ (the same parameter in \eqref{refplaneb}) and $p$ are two parameters both related to the spectral flow in 2D WCFT, and they parametrize the image plane $\cC_{img}$ and physical plane $\cC_p$ respectively. The parameter $\mathrm{t}$ is equal to the total shift in $\bar{y}$ direction (as we follow the closed contour around the twist operator $\s_{g_{r}}$ counter-clockwise on the replicated geometry $\S_{r}$) divided by $-2\pi i \mu$. The parameter $\kappa$ denotes the freedom we own to choose  the image plane $\cC_{img}$. In the following context we would like to fix the  image plane $\cC_{img}$ to be the reference plane with coordinates $(x,y)$. More precisely, we require that when  the total number of replica sheets $l$ equal to  $1$, we have 
\be  \bar x=x,\quad\Tilde{y}=\bar{y}=y, \quad \mathrm{when}~l=1.  \ee 
This requirement, combined with \eqref{trnas1}, gives the following constraint on the parameters at $l=1$
\be \quad \left( \kappa -\mathrm{t} \right) \lvert_{l=1}=0 \label{conorb} \ee 

Using the transformation rule \eqref{wcfttransf1}, we can get the left-hand side of \eqref{wcft-stress-charge}
\begin{align}
     \langle P(\bar{x}^i) \rangle_{\Sigma_{r}}&= \langle \frac{\partial \Tilde{x}}{\partial \bar{x} }\left( P(\Tilde{x})+\frac{k}{2} \frac{\partial \bar{y} }{\partial \Tilde{x}}  \right) \rangle_{\cC_{img}} \notag \\
     &=  \frac{l_0}{\bar{x}(\bar{x}-l_{0})} \frac{k}{2r} \left( \mathrm{t} - \kappa \right) \mu -\frac{kp}{2},\label{pcharge2} \\
     \langle T(\bar{x}^{i}) \rangle_{\Sigma_{r}}&= \langle \left( \frac{\partial \Tilde{x}}{\partial \bar{x}} \right)^{2} \left( T(\Tilde{x})-\frac{c}{12} \{f, \Tilde{x} \} \right)+ \frac{\partial \Tilde{x}}{\partial \bar{x}} \frac{\partial \Tilde{y} }{\partial \bar{x}} P(\Tilde{x})-\frac{k}{4}  \left(\frac{\partial \Tilde{y} }{\partial \bar{x}} \right)^{2}  \rangle_{\cC_{img}} \notag \\
     =& \frac{l_{0}^{2} }{\bar{x}^2(\bar{x}-l_0)^2} \frac{1}{r} \left( \frac{c}{24}\frac{r^2-1}{r }-\frac{k}{ 4r } \left( \mathrm{t}  -\kappa \right)^2 \mu^2 \right)   \notag \\
     & + p \frac{ k}{ 2r } \frac{l_0}{\bar{x}(\bar{x}-l_0)}  \left(\mathrm{t} -\kappa \right)\mu -\frac{k p^2}{4},   \label{tcharge2}
\end{align}
where in the final step of \eqref{pcharge2} and \eqref{tcharge2} we have used the fact that the image plane denoted by $\cC_{img}$ is chosen to be the reference plane, which have vanishing vacuum charges, $\langle P(\Tilde{x}) \rangle_{img}=0$ and $\langle T(\Tilde{x}) \rangle_{img}=0$.
By comparing \eqref{pcharge1}, \eqref{tcharge1} with \eqref{pcharge2} \eqref{tcharge2}, we obtain the conformal dimension and charge of the twist operator $\s_{g_r}$
\begin{align}
    h_{g_r} &= \frac{c}{24}\frac{r^2-1}{r^2}-\frac{k}{4r  } \left(  \mathrm{t} -\kappa \right)^2 \mu^2 = \frac{c}{24}\frac{r^2-1}{r}+\frac{q_{g_r}^{2}}{rk},  \\
    q_{g_r} &=i \frac{k}{2}  \left( \mathrm{t} -\kappa \right) \mu.  \label{qm}
\end{align} 
In the following we would use the spectral flow invariant conformal dimension $h^{inv}_{g_r}$ for convenience 
\be   h_{g_r}^{inv}= h_{g_r}-\frac{q_{g_r}^2}{rk}=\frac{c}{24}\frac{r^2-1}{r}  \label{hm}  \ee 
where  $rk$ appears in the denominator because the level related to the orbifold WCFT defined on $\S_{r}$ is $ r\times k $. 

Before we consider the general non-Abelian case, let  us first study a simple case: the twist operator $\s_{(12...r)}$ in the Abelian orbifold $\text{WCFT}^{\otimes r} / \mb{Z}^{r}$. In this case, as we follow the contour we have 
\be \bar{x}^{j}=x^{j} \to \bar{x}^{j}e^{2 \pi i}=x \, e^{2 \pi i} , \quad \bar{y}=y \to \bar{y}-2 \pi i \mu =  y-2 \pi i \mu ,\quad i=1...r  \label{twistmono} \ee 
so the total shift of $y$ direction is $-2 \pi i \mu r $, then $\mathrm{t}=r=l$. The condition in \eqref{conorb} tell us $\kappa=1$. We can see easily that the uniformization map \eqref{uniformap}
reduce to the one appearing in \cite{Chen:2019xpb}
\be \Tilde{x}=\left( \frac{\bar{x} }{\bar{x}-l_0} \right)^{\frac{1}{r}}, \quad  y -r\, \mu \log{\Tilde{x}}+ \mu \log{\Tilde{x}} =\bar{y}+p \bar{x},  \ee 
and reproduce the dimension $h_{r}^{inv}$ and charge $q_{r}$ of the twist operator $\s_{(12...r)}$
\be  h_{r}^{inv}=\frac{c}{24} \frac{r^2-1}{r}, \quad q_{r}=i \frac{k}{2}(r-1)\mu, \ee which are the same as the ones in \eqref{chargeRindler} obtained from the Rindler method. 



For a generic non-Abelian orbifold,  the monodromy conditions is tricky, as the field identifications induced by the twist operators are involved. To determine the monodromy condition of all twist operators in an orbifold WCFT, the guideline is to keep charge conservation. More explicitly, here is the guideline for the monodromy condition.

\textbf{Guideline:} \textit{The monodromy condition of twist operators in $\bar{x}$ direction is the same as the holomorphic coordinate $z$ of 2D CFT; For the $\bar{y}$ coordinate, when the shifts along $\bar{y}$ axis  at one branching point $(\bar{x}_0,\bar{y}_0)$ have been fixed by several independent twist operators $\s_{g_{i}}(\bar{x}_0,\bar{y}_0)$ as we go around $(\bar{x}_0,\bar{y}_0)$, 
then all the other twist operators which may appear as the anti-twist operators or in the OPE of the above twist operators can be determined completely.}

In the following, we present several non-trivial examples to show in detail how this guideline works. These examples appear in the computations of entanglement entropy, reflected entropy, entanglement negativity and odd entropy. 

\paragraph{Example 1:} \textbf{Anti-twist operator $\Tilde{\s}_{(12...l)}$} \\

This example appears in the calculation of single interval entanglement entropy $S(\cA)$. The relations of the sheets in the replicated geometry of Abelian orbifold WCFT have been fixed by the single operator $\s_{(12...l)}$ located at the left end-point of the interval $\cA$, which is just \eqref{twistmono}. Thus for the anti-twist operator $\Tilde{\s}_{(12...l)}$ located at the right end-point of $\cA$ should behave as follows: 
\be  \bar{x}^{j} \to \bar{x}^{j} e^{2 \pi i}, \quad \bar{y}^{j} \to \bar{y}^{j}+ 2 \pi i \mu, \quad j=1,2,...,r \label{yshif1}. \ee
Here we have identified the anti-twist operator $\tilde\sigma_{(12...l)}$ with $\sigma_{(l(l-1)...1)}$ inserted at $(l_0,0)$ on each sheet. Therefore, the route  is going around the branch point $(l_0,0)$ counterclockwise from the sheet $i$ to the sheet $i-1$. As a result, the phase acquired by $\bar x$ is the same as \eqref{twistmono} but the shift along $\bar y$ is opposite to \eqref{twistmono}. 
The total shift in $y$ direction is $2\pi i \mu r$. Then the parameters in \eqref{uniformap} have been determined to be
\be r=l, \quad \mathrm{t}=-l, \quad \kappa=-1 \label{anti1} \ee 
Then from \eqref{qm} and \eqref{hm} we get the dimension and charge of anti-twist operator
\be q=-i\frac{k}{2}(l-1)\mu, \quad h^{inv}=\frac{c}{24}\frac{l^2-1}{l} \ee 
which is consistent with charge conservation of correlation function $\langle \s_{(12...l)} \Tilde{\s}_{(12...l)} \rangle$. 

More generally, the anti-twist operator locating at the other end-point always has the same dimension but opposite charge with respect to a twist operator. This simply comes from the monodromy condition. 

\paragraph{Example 2:}\textbf{OPE of twist operators $\s_{(12...l)}\s_{(12...l)}$ with even $l$}  \\

\begin{figure}
    \centering
    \subfigure[]{
    \includegraphics[width=0.4\textwidth]{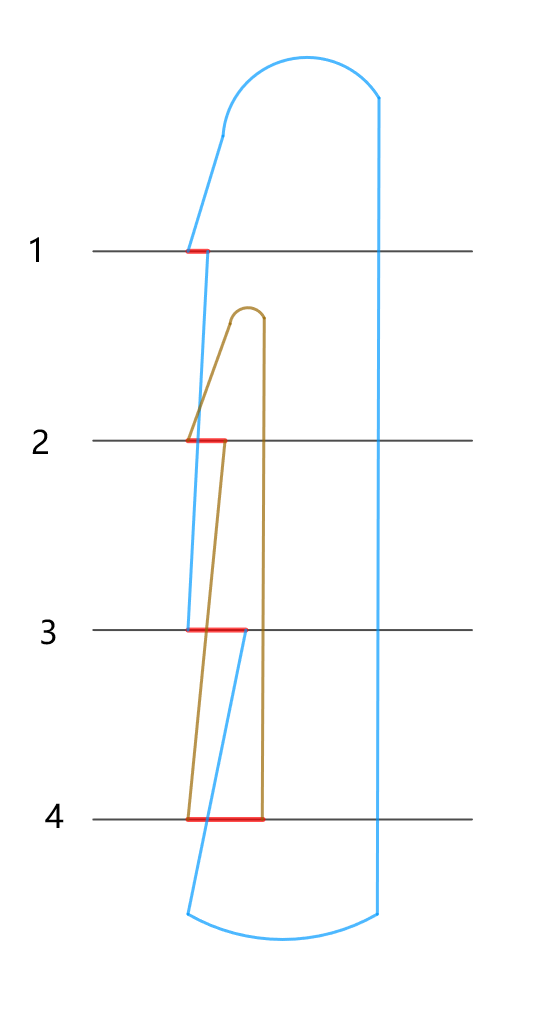}}
    \subfigure[]{
    \includegraphics[width=0.4\textwidth]{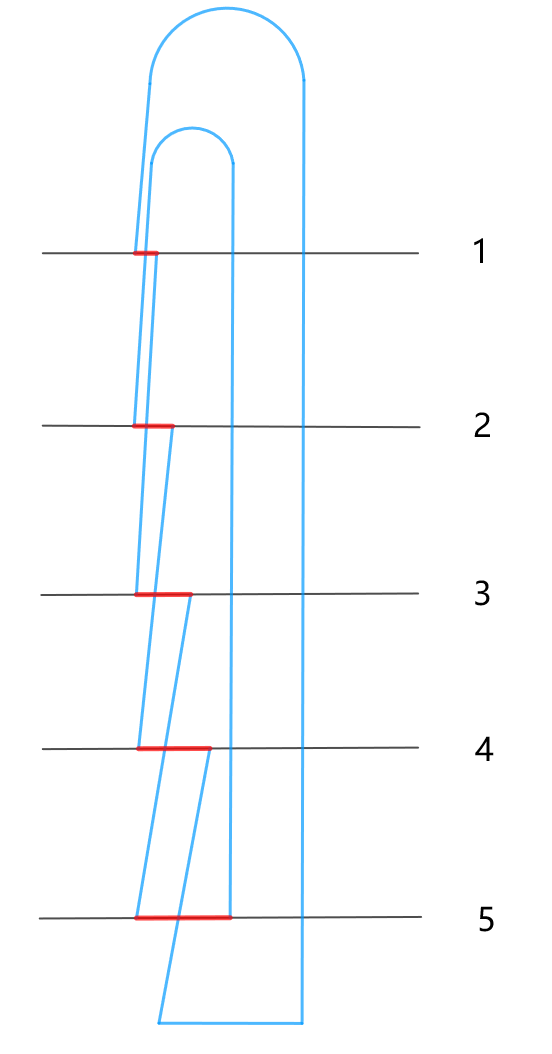}}
    \caption{Contour configurations representing the OPE of two twist operators $\sigma_{l}\sigma_{l}$. The red lines show the shifts in $y$ direction, which is fixed by $\sigma_{l}$ operator.  (a): A typical example for $l$ even ($l=4$) showing how the sheets decouples into two independent sets, each of $l/2$ replica sheets, and how the boundary condition in $y$ direction \eqref{2lbondarycond} are obeyed; (b):  A typical example for $l$ odd ($l=5$) showing the net effect is not just a reshuffling of the sheet numeration due to the non-trivial boundary condition in $y$ direction \eqref{example3}. }
    \label{fig:odd even ope}
    \end{figure}

This example would appear in the calculation of entanglement negativity $\cE(\r_{AB})$. When $l$ is an even number, the OPE of two same $l$-fold twist operators would generate two $l/2$-fold twist operators as the leading order contribution. For example when $l=4$, we have $\sigma_{(1234)}\sigma_{(1234)} \sim \sigma_{(13)(24)}  +...$, see Figure \ref{fig:odd even ope} (a). We view the twist operator $\sigma_{(13)(24)}$ as two independent twist operators $\sigma_{(13)}$ and $\sigma_{(24)}$ located at the same point to compute the quantum numbers of $\sigma_{(13)(24)}$ for clarity. The replicated geometry have been fixed by the single twist operator $\s_{(12...l)}$ for general $l$, then the twist operators $\s_{(13...l-1)}$ and $\s_{(24...l)}$ appearing in the OPE have the following monodromy conditions
\be \bar{x}^{j} \to \bar{x}^{j} e^{2 \pi i}, \quad \bar{y}^{j} \to \bar{y}^{j}-4 \pi i \mu, \quad j=1,3,...,l-1 \;\; \text{or} \;\;\;2,4,...,l.  \label{2lbondarycond}\ee
The total shift in $y$ direction is $-4\pi i \mu r$. This implies the parameters in \eqref{uniformap} for both twist operator $\s_{(13...l-1)}$ and $\s_{(24...l)}$  have the following relations
\be r=l/2, \quad \mathrm{t}=l, \quad \kappa=1, \label{anti2} \ee 
so from \eqref{qm} and \eqref{hm} we get their dimensions and charges 
\be h^{inv}=\frac{c}{24} \frac{(l/2)^2-1}{l/2}, \quad  q=i \frac{k}{2}(l-1)\mu.  \ee 
Then the total charge of twist operator $\sigma_{(13...l-1)(24...l)}$ are twice the above result, which is consistent with the charge conservation of the three-point correlation functions $\langle \sigma_{(12...l)}(\bar{x}_{1},\bar{y}_{1})$ $ \sigma_{(12...l)}(\bar{x}_{2},\bar{y}_{2})$ $\tilde\sigma_{(13...l)(24...l)} ((\bar{x}_{3},\bar{y}_{3}))\rangle$.\\

\paragraph{Example 3:}\textbf{OPE of twist operators $\s_{(12...l)}\s_{(12...l)}$ with odd $l$} \\

This example would appear in the calculation of the odd entropy $\cS_{o}(AB)$. When $l$ is an odd number, the OPE of two $l$-fold twist operators would generate another $l$-fold twist operator. For example when $l=5$, we have $\sigma_{(12345)}\sigma_{(12345)} \sim \sigma_{(13524)} +...$, see Figure \ref{fig:odd even ope} (b). Again the relations of $y$ direction of all the $l$ replica sheets have been fixed by the single twist operator $\s_{12...l}$, so for the twist operator $\s_{(13...(l)24...(l-1))}$ the monodromy conditions are 
\be \bar{x}^{j} \to \bar{x}^{j} e^{2 \pi i}, \quad \bar{y}^{j} \to \bar{y}^{j}-4 \pi i \mu, \quad j=1,2,...,l. \label{example3} \ee 
The total shift in $y$ coordinate now is $-4 \pi i \mu l$. Then the parameters are
\be r=l, \quad \mathrm{t}=2l, \quad \kappa=2  \label{anti3} \ee 
which tell us the dimension and charge for the twist operator $\s_{(13...(l)24...(l-1))}$ is 
\be h^{inv}=\frac{c}{24}\frac{l^2-1}{l}, \quad  q= i\frac{k}{2}(2l-2)\mu. \quad \ee
Note that due to the non-trivial monodromy conditions of the twist operator $\sigma_{(13...(l)24...(l-1))}$ in the $y$ direction, the net effect is not just a reshuffling of the sheet numeration. This is the key difference between the twist operators in orbifold CFT and orbifold WCFT. The result is consistent with the charge conservation of three-point correlation functions $\langle \s_{(12...l)}( \bar{x}_{1},\bar{y}_{1})$ $ \s_{(12...l)}( \bar{x}_{2}, \bar{y}_{2})$ $\Tilde{\s}_{(13...(l)24...(l-1))} ( \bar{x}_{3},\bar{y}_{3}) \rangle$.\\

\paragraph{Example 4:}\textbf{OPE of twist operators $\s_{g_{A}}\s_{g_{B}^{-1}}$ in the reflected entropy} \\

Let us consider a much more complicated example, the OPE of twist operator $\s_{g_{A}}\s_{g_{B}^{-1}}$ in \eqref{fusiontwist} in the non-abelian orbifold WCFT containing $mn$ replica sheets, which shows the power of our formulation. The independent twist operators determining the relations of all the $mn$ replicas can be chosen as $ \{ \s_{g_{A}},\s_{g_{B}} \} $ or $\{  \s_{g_{A}^{-1}},\s_{g_{B}}  \} $. See Figure \ref{fig:monodromy3} for more concrete showing, where the single twist operator $\s_{g_{B}}=$ $\s_{(1234)(5678)}$ determine the relations of all the replicas in the same $n$-column and the single twist operator $\s_{g_{A}}=$ $\s_{(1278)(5634)}$ allow us to determine the relations between replica sheets in different $n$-columns. In addition, we have 
\be  \s_{(1278)(5634)}   \Tilde{\s}_{(1234)(5678)}=  \s_{(26)(48)}+..., \quad g_{B}^{-1}g_{A}=(\tau_{k}^{0})^{-1} \tau_{k}^{m/2} \ee
where $(\tau_{k}^{0})^{-1}$ and $\tau_{k}^{m/2}$ are two $n$-cyclic permutation group elements appearing in \eqref{gaga}. As in \textbf{Example 2}, to compute the dimension of twist operator $\s_{(26)(48)}$, we separate it into two twist operators $\s_{(26)}$ and $\s_{(48)}$ located at the same point. Closely following the contours in Figure \ref{fig:monodromy3}, we would get a conclusion that as we go around counter-clockwise by a $2\pi$ angle from $j$-th replica sheet to $k$-th replica sheet with 
\be (j,k)=(1,5) \;  \text{or} \; (2,6) \; \text{or}\; (3,7) \; \text{or} \; (4,8) \ee 
we have $\bar{y}^{j}\to \bar{y}^{k}$ with no shift, which means the total shift in $y$ direction vanishes. Then we have the parameters for both twist operators $\s_{(\tau_{k}^{0})^{-1}}$ and $\s_{\tau_{k}^{m/2}}$ for general $m$ and $n$
\be r=n, \quad \mathrm{t}=0, \quad \kappa=0, \quad l=m\times n.  \label{anti4} \ee 
These give us the dimension and charge of $\s_{g_{B}^{-1}g_{A}}$
\be  h^{inv}=2 \frac{c}{24} \frac{n^2-1}{n}, \quad q=0  \ee 
This case is a true non-abelian orbifold calculation, and the $l=1$ condition is equivalent to $m=n=1$.


\begin{figure}
    \centering
    \includegraphics[width=0.3\textwidth]{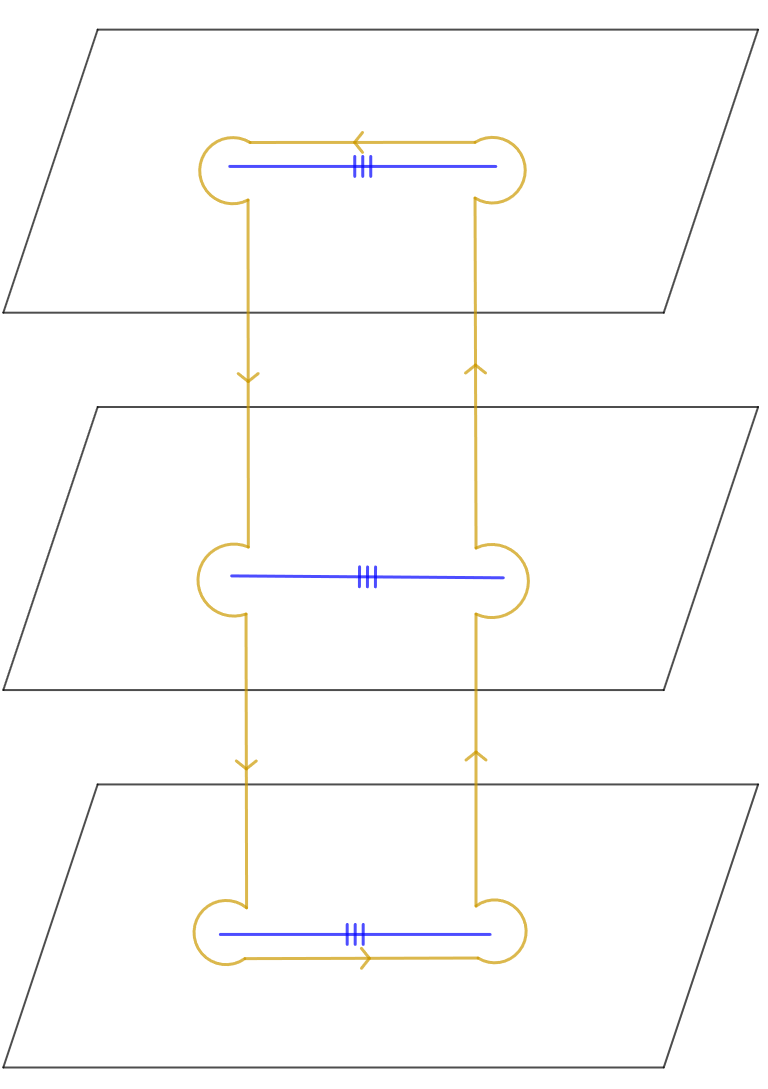}
    \caption{A (brown) closed contour without touching the (blue) branching cuts representing the boundary interval $\cA$ in the abelian orbifold $\S_{3}$ for computing entanglement entropy. The total shift in $y$ direction is zero if we go around the contour counter-clockwise following the arrows.}
    \label{fig:contour}
\end{figure}

There is a nice interpretation of the guideline on the above monodromy conditions. It implies that any closed contour not crossing any branch cut in the replicated geometry $\S_{l}$  has a total zero shift in $y$ direction. Such configuration is relevant to the computation of the correlation function of the twist operators $\s_{g_{i}}$, whose related group elements satisfy 
\be \prod_{i} g_{i}= \mathbf{1}.   \ee
In order that the correlation function is non-vanishing, the charges must be conserved. The charge conservation then requires that the total amount of the shifts in $\bar{y}$ direction must be vanishing.  See Figure \ref{fig:contour} for an explicit example. 

Technically, as shown in the above examples,  the amount of shifts in $\tilde y$, the U(1) direction of the image place, is not always $-2\pi i\mu$,  as what we have for the twist operator $\sigma_{(12...l)}$ when computing the entanglement entropy. Instead, the amount of shifts varies case by case. This can be seen immediately from the uniformization map  \eqref{uniformap}. Due to the definition of $\mathrm{t}$, we can see that $\tilde y$ is shifted by $2\pi \kappa i\mu$, i.e. $\kappa$ times the periodicity of the reference plane, with $\kappa$ taking different values in specific examples.  This suggests that when $\bar y$ is shifted by $-2\pi i\mathrm{t}\mu$ following the contour suggested by the twist operator, the image of this contour encircles the $\tilde y$ direction by $\kappa$ times. 

\subsection{Reflected entropy in 2D holographic WCFT}

In this subsection, let us compute the reflected entropy $S_{R}(A:B)$ in holographic 2D WCFT. The results can be summarized as follows:
\begin{itemize}
    \item The reflected entropy $S_{R}(A:B)$ of two disjoint intervals $A$, $B$ in WCFT is independent of the $U(1)$ direction,  similar to the case of mutual information $I(A:B)$. This means that it is a spectral-flow invariant quantity with the same value on any spectral flowed plane for the same entangling configuration.
    
     \item  For the same entangling configuration of two disjoint intervals $A$, $B$ on both the vacuum state and thermal state, we have the following relation between WCFT and CFT,
    \be S_{R}^{WCFT}(A:B)=\frac{1}{2} S_{R}^{CFT } (A:B).\ee 
 
\end{itemize}
We now show the computational process to reach the above statements. From the study in the above subsection, we can read easily the conformal dimensions $h^{inv}$ and $U(1)$ charges $q$ of various twist operators in the relevant orbifold WCFT. Since we are dealing with WCFT on a replicated  geometry of $n m$ sheets, both $c$ and $k$ need to be multiplied by the number of copies $n m$. Therefore, the relation between $h$ and $h^{inv}$ for twist operators is $h^{inv}=h-\frac{q^2}{n m k}$, where $k$ is the level of the original WCFT defined on a single sheet. In the end, we get the following results. 

\begin{itemize}
    \item $\sigma_{g_{A}}$, $\sigma_{g_{B}}$ : 
    \be
    \begin{aligned}
   & h_{\sigma_{g_A}}^{inv}=h_{\sigma_{g_B}}^{inv}=nh_{m}^{inv}=\frac{nc}{24}(m-\frac{1}{m}), \\ &q_{\sigma_{g_A}}=q_{\sigma_{g_B}}=nq_{m}=n(m-1)\frac{ i\mu k}{2}.
    \end{aligned}\ee
    
     \item $\sigma_{g_{A}^{-1}}$, $\sigma_{g_{B}^{-1}}$ : 
     \be
    \begin{aligned}
   & h_{\sigma_{g^{-1}_A}}^{inv}=h_{\sigma_{g^{-1}_B}}^{inv}=nh_{m}^{inv}=\frac{nc}{24}(m-\frac{1}{m}), \\ &q_{\sigma_{g^{-1}_A}}=q_{\sigma_{g^{-1}_B}}=-nq_{m}=-n(m-1)\frac{ i\mu k}{2}
    \end{aligned}\ee
    \item $\sigma_{g_{A}g_{B}^{-1}}$,
    \be\label{h-q-gagb} h_{g_{A}g_{B}^{-1}}^{inv}=2h_{n}, \quad q_{g_{A}g_{B}^{-1}}=0 .\ee 
\end{itemize}

For the important OPE coefficients, we have 
\be \label{wcft-ope}\sigma_{g_{A}^{-1}} \sigma_{g_{B}} = C_{n,m} \sigma_{g_{B} g_{A}^{-1}}+...\quad , \quad  e^{\pi\mu q_n(n-1)}C_{n,m}=(2m)^{-2h_{n}}. \ee 
 This relation can be obtained by using the same method as in the CFT case. We choose to compute this particular three-point function on the reference plane $\cC_{ref}$, which corresponds to choosing $p=0$ in \eqref{uniformap} for the original physical plane to do the replication.  With this choice, the coordinate $\bar{x}$ for $\cC_p$ becomes $x$ for the reference plane hereafter.
The image plane can be chosen to be the reference plane without loosing generality. Then we find 
\begin{align}\label{wcft-ope1}
    & \langle \sigma_{g_{A}^{-1}} (x_{1},y_{1}) \sigma_{g_{B}}(x_{2},y_{2}) \sigma_{g_{A} g_{B}^{-1}}(x_{3},y_{3}) \rangle_{\mbox{\small WCFT}^{\otimes nm}(ref)} \nn \\
   =& e^{S_{L}(\phi)}\underbrace{\Big\lvert\frac{\partial s^{+}}{\partial x}\Big\lvert^{2h_{n}}_{s=s_{1}^{+}} \Big\lvert\frac{\partial s^{+}}{\partial x}\Big\lvert^{2h_{n}}_{s=s_{2}^{+}} 
     \langle \sigma_{(\tau_{n}^{0})^{-1}}(s_{1}^{+},s_{1}^{-}) \sigma_{\tau_{n}^{m/2}}(s_{2}^{+},s_{2}^{-}) \rangle_{\mbox{\small WCFT}^{\otimes n}(ref)}}_{\left( \cD \right)}  \nn\\
     =& \left. \left( \langle \sigma_{g_{A}^{-1}} (x_{1},y_{1}) \sigma_{g_{B}}(x_{2},y_{2}) \rangle_{\mbox{\small WCFT}^{\otimes m}(ref)} \right)^{n} \right|_{A=B} \times \left( \cD \right) \nn\\
     =& e^{-\pi\mu q_n(n-1)}e^{inq_{m}(y_1-y_2) } (2m)^{-2h_{n}}|x_{32}|^{-2h_{n}}|x_{31}|^{-2h_{n}}|x_{12}|^{-2nh_{m}+2h_{n}}.
   \end{align}
The first equality follows from a warped conformal transformation
\be s^+=\frac{(x-x_{1})^{1/m}}{(x-x_{2})^{1/m}}, \hs{3ex} s^--(n-1)\mu \log{s^{+}} =y, \ee
which maps the replicated manifold composed of $n m$ reference planes to a replicated manifold composed of $n$ reference planes. In addition, the twist operators $\sigma_{(\tau_{n}^{0})^{-1}}(s_1^+,s_1^-) $ and  $\sigma_{\tau_{n}^{m/2}}(s_2^+,s_2^-)$ have the same conformal dimensions $h_{n}$ but opposite charges $q_n$ and $-q_n$, with the following coordinates
\bea 
s^+_1=\frac{(x_3-x_{1})^{1/m}}{(x_3-x_{2})^{1/m}}, && s^-_1-(n-1)\mu \log{s^{+}_1} =y_3, \label{s1}\\
s^+_2=e^{i\pi}\frac{(x_3-x_{1})^{1/m}}{(x_3-x_{2})^{1/m}}, && s^-_2-(n-1)\mu \log{s^{+}_2} =y_3.\label{s2}
\eea 
From \eqref{wcft-ope1}, we can easily read off the OPE coefficient $C_{n,m}$ as well as the conformal dimension and the charge of $\sigma_{g_Ag_B^{-1}}$ as claimed in \eqref{h-q-gagb} and \eqref{wcft-ope}. Compared to the CFT case, there is an additional factor $e^{-\pi\mu q_n(n-1)}$  coming from the correlation functions of $\sigma_\tau$'s with the coordinates \eqref{s1} and \eqref{s2} and the fact that twist operator $\sigma_{(\tau_{n}^{0})^{-1}} $ and $\sigma_{(\tau_n^{m/2})}$ have charge $-q_n$ and $q_{n}$ respectively. More precisely there is
\begin{equation}
    \langle \sigma_\tau\sigma_\tau\rangle\sim e^{iq_n(s_{1}^- -s_{2}^-)}=e^{iq_n\mu(n-1)\log({s^+_1}/{s^+_2})} =e^{-\pi\mu q_n(n-1)} \nn
\end{equation}
However this factor would not affect  the reflected entropy, since its exponent scales like $(n-1)^2$ that would vanish in the $n\to 1$ limit. Apart from this extra factor, the OPE coefficient is actually the square root of the one in the CFT case. This is due to the fact that we have only  the uniformization map on $x$, half of the ones in CFT.\\


Now we are ready for the computation of the reflected entropy in the holographic WCFT. Let us first consider the reflected entropy in the vacuum state. We need to compute the four-point function of the twist operators on the plane
\be
\langle \sigma_{{g}_A}(x_1,y_1) \sigma_{{g}^{-1}_A}(x_2,y_2) \sigma_{{g}_B}(x_3,y_3) \sigma_{{g}^{-1}_B} (x_4,y_4) \rangle_{\mbox{\small WCFT}^{\otimes n m}({\cal C}_{p})} 
\ee 
As in the CFT case, we care about the $t$-channel expansion, which is equivalent to take the limit of cross ratio $\mathrm{x}\to 1$. 
One essential assumption is the dominance of a single block in the large $c$ limit of holographic WCFT. This allows us to find  
\begin{align}
    & \langle \sigma_{{g}_A}(x_1,y_1) \sigma_{{g}^{-1}_A}(x_2,y_2) \sigma_{{g}_B}(x_3,y_3) \sigma_{{g}^{-1}_B} (x_4,y_4) \rangle_{\mbox{\small WCFT}^{\otimes n m}(\cal C_{\mu})}  \notag \\
    =& \frac{e^{i q_{\sigma_{g_A}}(y_{12}+y_{34}-p(x_{12}-x_{34})) }}{x_{41}^{2h_{\sigma_{g_A}}}x_{32}^{2h_{\sigma_{g_A}}}} \sum_{p} C_{p12}C^{p}_{34} (1-\mathrm{x})^{\frac{2q_{\sigma_{g_A}}^{2}}{mnk}} \mathrm{x}^{-\frac{2 q_{\sigma_{g_A}}^{2}}{mnk}} \mathcal{F}(mnc-1,h_{i}^{inv},h_{p}^{inv},1-\mathrm{x}) \notag \\
    & \approx \underbrace{\frac{e^{i n q_{m} (y_{12}+y_{34}-p(x_{12}-x_{34})) }}{x_{41}^{ 4nh_{m}}x_{32}^{4nh_{m}}}(1-\mathrm{x})^{ \frac{ 2(n q_{m})^{2}}{mnk} } \mathrm{x}^{-\frac{2(n q_{m})^{2}}{mnk}}}_{\big(\cal D \big)} (C_{n,m})^2   e^{- (2h_{n})  \ln{\frac{ 1+\sqrt{\mathrm{x}} }{ 4(1-\sqrt{\mathrm{x}}) }}}. \label{wcft32}
\end{align}
 Note that in the $\cal D$ part of expression \eqref{wcft32},  the exponent of each term is proportional to $n$. It is exactly this fact that makes the $\cal D$ part cancel out between the numerator and the denominator in evaluating the reflected entropy. Apart from the $\cal D$, the form of the rest part of \eqref{wcft32} is similar to the  square root of the contribution part of CFT in \eqref{4pttwistt}, which leads us to the final result
\be \label{cft-wcft-pl} S^{ WCFT }_{R;vac}(A:B) =\frac{1}{2} S^{ CFT }_{R;vac}(A:B). \ee


Next, we consider the refleted entropy in thermal state. Now the four-point function is defined on  the thermal cylinder $(w^{+},w^{-})\sim (w^{+}+i \beta,w^{-}-i \bar{\beta} )$. From the transformation property \eqref{npt-trans} of primary operators, we have the relation: 
 \begin{small}
 \begin{align}
     & \langle \sigma_{{g}_A}(w_{1}^{+},w_{1}^{-}) \sigma_{{g}^{-1}_A}(w_{2}^{+},w_{2}^{-}) \sigma_{{g}_B}(w_{3}^{+},w_{3}^{-}) \sigma_{{g}^{-1}_B} (w_{4}^{+},w_{4}^{-}) \rangle _{\mbox{\small WCFT}^{\otimes n m}}^{cylinder} \notag \\
         =&|x'(w_{1}^{+})x'(w_{2}^{+})x'(w_{3}^{+})x'(w_{4}^{+})|^{h_{g_{A}}} \langle \sigma_{{g}_A}(x_{1},y_{1}) \sigma_{{g}^{-1}_A}(x_{2},y_{2}) \sigma_{{g}_B}(x_{3},y_{3}) \sigma_{{g}^{-1}_B} (x_{4},y_{4}) \rangle _{\mbox{\small WCFT}^{\otimes n m}}^{ref;plane}.
\end{align}
 \end{small}
 A straightforward computation leads to 
 \be\label{cft-wcft-thermal}
 S^{ WCFT }_{R;thermal}(A:B)=\frac{1}{2} S^{ CFT }_{R;thermal}(A:B).  \ee
As in the holographic CFT, and the  $s$-channel expansion at the leading order of large $c$ gives 
\be S^{WCFT}_{R;vac}(A:B)=S^{WCFT}_{R;thermal}(A:B)=0. \label{eq22} \ee 
Consequently, we may expect a first-order phase transition when the cross ratio increases from $0$ to $1$. 

\section{Pre-Entanglement Wedge and AdS$_3$/WCFT}

In this section we would identify the bulk geometric dual of boundary reflected entropy in the AdS$_3$/WCFT correspondence, by using the swing surface proposal described in \cite{Apolo:2020qjm,Apolo:2020bld}.  A crucial step is to find which bulk region in AdS$_3$/WCFT  is the natural analogue of the entanglement wedge in AdS/CFT.  In what follows, we briefly review the holographic correspondence between the AdS$_3$ gravity with the Comp\`ere-Song-Strominger (CSS) boundary conditions and 2D WCFT. 


In three dimensions, a set of consistent asymptotically boundary conditions is essential to define a  gravity theory. The transformations keeping these boundary conditions form an asymptotic symmetry group(ASG), which  restrict the boundary field theory dual. In AdS$_3$ gravity, under the well-known Brown-Henneaux boundary conditions\cite{Brown:1986nw}, the ASG is generated by two copies of the Virasoro algebra, suggesting a CFT dual.  In contrast, under the CSS boundary conditions,  the asymptotic symmetry group (ASG) is generated by a left-moving Virasoro and a $U(1)$ Kac-Moody algebra \cite{Compere:2013bya}, leading to the so-called AdS$_3$/WCFT correspondence. The similar  Virasoro-Kac-Moody algebra generates the ASG of 3D topological massive AdS$_3$ gravity  under appropriate boundary conditions as well\cite{Anninos:2008fx,Compere:2009zj}, leading to the (W)AdS$_3$/WCFT correspondence. Here we focus on the AdS$_3$/WCFT correspondence.

Denoting the lightcone coordinates of global AdS$_3$ on a cylinder by 
 \be \left( u=\phi + t, \; v=\phi-t, \; r \right), \quad \phi\sim \phi+2 \pi, \ee
then the CSS boundary conditions are as follows 
\begin{align}
    g_{rr}=\frac{1}{r^2}+ \cO(\frac{1}{r^4})&, \quad  g_{ru}=\cO(\frac{1}{r^3}), \quad g_{rv}=\cO(\frac{1}{r^3}), \quad g_{uv}=\frac{r^2}{2}+\cO(1),  \notag\\
    &g_{uu}=r^2 J'(u)+\cO(1), \quad g_{vv}=T_{v}^2 + \cO(\frac{1}{r}), \label{css}
\end{align} 
where $J(u)$ is an arbitrary function and
\be 
c=\frac{3 }{2G},
\ee 
with $G$ being the 3D Newton's constant. Here we have set the AdS scale to be $1$, as before. The parameter $c$ appears as the central charge in the dual WCFT.   In the Fefferman-Graham gauge, \eqref{css} is sufficient to determine the phase space of solutions in pure Einstein AdS$_3$ gravity in terms of two functions $L(u)$, $J(u)$ and a constant $T_{v}$,
\begin{align}
    ds^2=\frac{dr^2}{r^2}&+r^2 du\left[ dv+J'(u)du \right]+ L(u) du^2+T_{v}^2 \left[ dv+J'(u) du\right]^2 \notag \\
    &+\frac{1}{r^2} T_{v}^2 L(u)du \left[dv+J'(u)du \right]. \label{phasecss}
\end{align} 

We will mainly work in the zero mode backgrounds with $L(u)=T_{u}^{2}$ and $J(u)$ being constants. For these backgrounds, at the $ r\to \infty$ conformal boundary, the coordinates $(u,v)$ are related to the coordinates $(\hat{x}',\hat{y}')$ of the canonical cylinder  by a state-dependent map \cite{Detournay:2012pc} 
\be u=\hat{x}', \quad v=\hat{x}' +\frac{\hat{y}'}{2 \mu T_{v} } ,\label{bbmap22} \ee 
where $\mu$ is a parameter characterizing the WCFT given by \eqref{match12}. In these cases, \eqref{phasecss} describes a subset of the space of solutions of three dimensional gravity that are also compatible with Brown-Henneaux boundary conditions, and the energy and angular momentum associated with such backgrounds are respectively given by $M=(T_{u}^{2}+T_{v}^{2})/4G$ and $J=(T_{u}^{2}-T_{v}^{2})/4G$. These zero mode solutions include the the global $AdS_3$ vacuum with $M=-1/8G$ and $J=0$, the conical defect geometries with $ -1/8G<M<0$, $-1/8G<J<1/8G$, and the BTZ black holes with $M\ge |J|\ge0 $. When $L(u)$ or $J'(u)$ are not constants, the corresponding backgrounds are interpreted as the solutions dressed with additional boundary gravitons.


\subsection{Approximate modular flow and the swing surface}

In \cite{Apolo:2020bld}, it was proposed that the holographic entanglement entropy in the AdS$_3$/WCFT and Flat$_3$/BMSFT correspondences could be described by a swing surface. In this subsection, let us review this proposal briefly. 

For the case where is an exact boundary modular flow generator $\zeta$, the assumptions that the Killing vectors of bulk vacuum state corresponds to the symmetry generators of WCFT vacuum state make us have the ability to extend the boundary modular flow generator into the bulk spacetime with no obstacle. However, for more general field states the modular Hamiltonian is generally nonlocal, and the modular flow generator can no longer be expressed as the linear combinations of boundary symmetry generators $h_{i}$ or the bulk Killing vector fields $H_{i}$ which may not exist. In \cite{Apolo:2020bld} the authors generalized the approach in \cite{Lashkari:2016idm} by constructing the modular flow near the swing surface in AdS$_3$/WCFT holography. The key steps of their construction can be summarized as follows, 
\begin{itemize}
    \item Find approximate boundary modular flow generator $\zeta^{(p)}=\sum_{i} a_{i}^{(p)} h_{i}=2 \pi \partial_{\tau_{(p)}}$
    around the endpoints of the subregion $p \in \partial \cA $. For 2D WCFT, this can be obtained simply from the ones for the vacuum state by sending the other endpoint to infinity.
    \item Extend the $d$-dimensional vector field $\zeta^{(p)}$ into a $(d+1)$-dimensional vector field $\xi^{(p)}_{\infty}$ near the asymptotic boundary,
    \be \xi^{(p)}_{\infty}=\sum_{i} a_{i}^{(p)} H_{i}, \quad \text{with} \;\; H_{i}|_{\partial \cM}=h_{i}, \;\; \xi^{(p)}_{\infty}|_{\partial \cM}= \zeta^{(p)} ,\label{bbmatch1}  \ee 
    where $H_{i}$ are the approximate Killing vectors near $p\in \cA$.
    
    \item Extend the $(d+1)$-dimensional vector field $\xi^{(p)}_{\infty}$ defined near the asymptotic boundary into the bulk following the null geodesics $\gamma_{(p)}$, which defines a null vector field $\xi^{(p)}$ along $\gamma_{(p)}$ satisfying 
    \be \xi^{(p) \mu} \xi^{(p)}_{\mu}=0, \quad \xi^{(p)}|_{\partial \cM}=\zeta^{(p)}, \quad \xi^{(p) \mu} \nabla_{\mu} \xi^{(p)\nu}=\pm 2\pi \xi^{(p)\nu}.  \label{bbmatch2} \ee 
    
    \item Extend the vector field $\xi^{(p)}$ defined on the null geodesics $\gamma_{(p)}$ into the vector field on the whole $(d+1)$-dimensional spacetime by requiring that 
   \be \xi|_{\gamma_{(p)}}=\xi^{(p)}, \quad \xi|_{\gamma_{\xi}}=0, \quad \nabla_{\mu} \xi^{\nu}|_{\gamma_{\xi}}=2 \pi n^{\mu \nu}, \quad \left( \cL_{\xi} g \right)_{\mu\nu} \sim \cO(r^{-\alpha_{\mu\nu }}) \; as \; r \to \infty \ee 
   where $\gamma_{\xi}$ is the set of fixed points of $\xi$ and $\alpha_{\mu\nu }$ are constant. These conditions make sure that $\xi$ is a correctly constructed asymptotic Killing vector at conformal boundary $r \to \infty$ and behaves like the boost generator in the local Rindler frame near $\gamma_{\xi}$. 
 \end{itemize}

With the above constructions we can identify the swing surface which gives the holographic  entanglement entropy. The swing surface is the union of the ropes and the bench
\be  \g_{\cA}=\g \cup \g_{b \p}, \quad \text{where} \quad  \g_{b \p}\equiv \cup_{p\in \p \cA} \g_{(p)},    \ee 
where $\g_{b\p}$ are called the $\textit{ropes}$ of the swing surface, which is a collection of null geodesics tangent to the approximate bulk modular flow around the point $p$ at the conformal boundary, and $\g$ is called the $\textit{bench}$ of the swing surface, which are subset of the fixed points of the constructed approximated bulk modular flow bounded by the null ropes. Actually, there exist another equivalent prescription for this swing surface which is very similar to the construction of HRT surface in the standard AdS/CFT case: the swing surface \cite{Apolo:2020bld} of single boundary interval $\cA$ is the extremal surface, which is the one with minimal area, that bounded by $\g_{b \p}$.  Then the holographic entanglement entropy is finally given by 
\be \label{swing-prop}
   S_\cA=\min \underset{\g_\cA\sim \cA}{\mathrm{ext}}\frac{\mbox{Area}(\g_\cA)}{4G}.
\ee

\subsection{Pre-entanglement wedge in AdS$_3$/WCFT}

In this subsection we try to find the analogue of entanglement wedge and causal wedge in  AdS$_3$/WCFT. During this process we would also like to emphasize some geometric properties of the swing surface and discuss the phenomena related to single interval entanglement phase transition. 


 Now it is not possible to simplify the calculation by choosing constant time slice respecting time reversal symmetry like the case in AdS/CFT although the bulk spacetime is static. We choose to work in the following gauge in 3D Lorentzian spacetime directly. The line element takes the form
\be \label{wcft-gauge}
ds^2= \frac{d \rho^2}{4(\rho^2-4T_{u}^2 T_{v}^2)}+ \rho \, du dv+ T_{u}^2 du^2+T_{v}^2 dv^2,
\ee 
which is related to the Fefferman-Graham gauge by a redefinition of the radial coordinate $\rho=r^2+ T_{u}^2 T_{v}^2/r^2$, and to the Boyer-Lindquist-like coordinates $ (t,r,\phi) $ by the following coordinate transformations  
\be   \rho=r^2-4M, \quad u=\phi+t, \quad v=\phi-t   .\label{BLcoordinate} \ee  
The boundary interval $\cA$ is parametrized by the bulk coordinates \eqref{wcft-gauge} as 
\be \label{interval-A}
\partial \cA=\{(u_{-},v_{-}),(u_{+},v_{+})\}, \quad u_{+}-u_{-}=l_{u}, \quad v_{+}-v_{-}=l_{v} ,\ee
which is related to the boundary parametrization by the bulk-boundary  map \eqref{bbmap22}.

The parametrization of the swing surfaces can be described as follows. The determination of null ropes turns out to be the most crucial step in the entire construction because it enables the boundary information to come in. With the help of two conservation laws related to two commuting Killing vectors $\p_{u}$, $\p_{v}$ of the zero-mode background \eqref{wcft-gauge} and the null geodesic condition, the two ropes emitted from the boundary endpoints should satisfy 
\be T_{u}^2 u'+ \frac{\rho v'}{2}=p_{u}, \quad  T_{v}^2 v'+ \frac{\rho u'}{2}=-p_{v}, \quad \frac{\rho'^2}{4(\rho^2-4T_{u}^2 T_{v}^2)}+T_{u}^2 u'^2+T_{v}^2 v'^2+\rho u' v'=0 ,\ee 
where the null ropes $\left(\r(\l), u(\l),v(\l)  \right)$ are parametrized by the affine parameter $\l$, and $p_{\mu}$, $p_{\nu}$ denote the momenta along the $u$, $v$ directions. Rewriting the above equations as 
\be \rho'=4 \sqrt{\rho p_{u} p_{v}+T_{u}^2 p_{v}^{2}+T_{v}^2 p_{u}^{2}}, \quad u'= -\frac{4 T_{v}^2 p_{u}+2\rho p_{v}}{\rho^2-4 T_{u}^2 T_{v}^2}, \quad   v'=\frac{4 T_{u}^2 p_{v}+2\rho p_{u}}{\rho^2-4 T_{u}^2 T_{v}^2}, \label{null-geo-eqn} \ee
we can see that if we require the ropes to reach the asymptotic boundary $\rho\to\infty$ at large $\lambda >0$, then the momenta $p_{u}$ and $p_{v}$ should satisfy $p_up_v\geq0$. For the case of $p_up_v>0$, the solutions to \eqref{null-geo-eqn} are given by
\begin{align}
   & \rho = \frac{p_{u}^{2} p_{v}^{2}(2 \lambda+\lambda_{0})^2-p_{u}^{2}T_{v}^2-p_{v}^{2}T_{u}^2 }{p_{u} p_{v}}, \nn \\
   & u =-\frac{1}{4T_{u}} \log{\{\frac{p_{v}^{2}[T_{u}+p_{u} (2 \lambda+\lambda_{0})^2 ]^2-p_{u}^{2}T_{v}^2  }{p_{v}^{2}[T_{u}-p_{u} (2 \lambda+\lambda_{0})^2 ]^2-p_{u}^{2}T_{v}^2} \}} +u_{0},  \label{solurope2} \\
   & v = \frac{1}{4T_{v}} \log{\{\frac{p_{u}^{2}[T_{v}+p_{v} (2 \lambda+\lambda_{0})^2 ]^2-p_{v}^{2}T_{u}^2  }{p_{u}^{2}[T_{v}-p_{v} (2 \lambda+\lambda_{0})^2 ]^2-p_{v}^{2}T_{u}^2} \}} +v_{0}, \nn  
\end{align}
where $\lambda_0$, $u_0$ and $v_0$ are integration constants. There are two remarkable points: 1) $p_{u}=0$ or $p_{v}=0$ is the singular limit of the solutions \eqref{solurope2}, which means that we should solve them directly from \eqref{null-geo-eqn}; 2) For $p_{u} \ne 0$ and $p_{v} \ne 0$, the above solutions are symmetric about $u$ and $v$ coordinates (up to an irrelevant minus sign), and this symmetry is not broken at the cutoff surface $\r=\r_{\infty}$. The second one would force the projection of tangent vectors of the null geodesics on the asymptotic boundary to  the symmetric one, which is the dilatation generator. However in a WCFT, the modular flow is not symmetric about $u$ and $v$ directions, thus the possible solutions of the swing surface can only exist in the singular limit of \eqref{solurope2}, i.e., $p_{u}=0$ or $p_{v}=0$ (but not both). It has been shown in \cite{Apolo:2020bld} that when choosing $T_{v}^2 >0$, the null ropes are the ones with $p_{v}=0$. If we denote the future directed one as $\g_{+}$, then it has momentum $p_{u}<0$. Similarly, the past directed one has $p_{u}>0$. 

Setting $p_{v}=0$ in \eqref{null-geo-eqn}, we get 
\be \rho'=4 \sqrt{T_{v}^2 p_{u}^{2}}, \quad u'= -\frac{4 T_{v}^2 p_{u} }{\rho^2-4 T_{u}^2 T_{v}^2}, \quad  v'=\frac{ 2\rho p_{u}}{\rho^2-4 T_{u}^2 T_{v}^2}, \label{nullropes}  \ee 
which have the solutions
\begin{small}
\be \gamma_{\pm}: \quad \rho =4|p_{u}|T_{v}\lambda +\rho_{0}, \;\;  u=\mp \frac{1}{4T_{u}}\log(\frac{\rho+2T_u T_v}{\rho-2T_u T_v})+{u}_{\pm}, \;\; v=\mp \frac{1}{4T_{u}}\log(\frac{\rho^2-4T_u^2 T_v^2}{\rho_\infty^2 })+{v}_{\pm}. \label{nullsolu12} \ee 
\end{small}
In the above solution, $\r_{0}$ is the integration constant, $u_{\pm}$, $v_{\pm}$ are the coordinates of boundary interval endpoints \eqref{interval-A} that are determined by the  bulk-boundary matching condition, and $\r_{\infty}$ denotes the radial location of the cutoff surface.  The null ropes have the following properties which can be seen easily from \eqref{nullropes} and \eqref{nullsolu12}:
\begin{itemize}
    
    \item For fixed zero-mode background \eqref{wcft-gauge} with constant $T_{u}$ and $T_{v}$, the rate of change of null ropes $\frac{d \rho}{d u}$ and $\frac{d \rho}{d v}$ related to any single boundary interval are always the same, which can be seen clearly from \eqref{nullropes}. 
    
    \item As we go toward the boundary, i.e., increase the value of $\rho$, the radial projection of the slope $|du/dv|$ of null ropes on the cut-off surface $\rho=\rho_{\infty}$ becomes smaller
    \be \frac{du}{dv}=-\frac{2T_{v}^{2}}{\rho}<0.\ee
  In addition, at each endpoint of the boundary interval, there are two null geodesics with $p_{v}=0$ and opposite values of $p_{u}$. However, only one of them can be the candidate of the ropes in the construction of swing surface,  due to the matching condition \eqref{bbmatch1} and \eqref{bbmatch2}. More precisely, the future outgoing lightray has $p_{u}<0$ and past outgoing lightray has $p_{u}>0$, which can be checked by
    \be dt=du-dv=-\frac{2 p_{u}(2T_{v}^{2}+\rho)}{\rho^2-4 T_{u}^{2} T_{v}^{2}} d\lambda  \propto -p_{u} .\ee 
    
    \item The condition $p_{v}=0$ of the null ropes ensures that they are perpendicular to both the boundary of the domain of dependence of the boundary interval and the bench of the swing surface, which all purely extend along the $v$ direction. 
    \end{itemize}

    \begin{figure}
        \centering
        \includegraphics[width=0.43\textwidth]{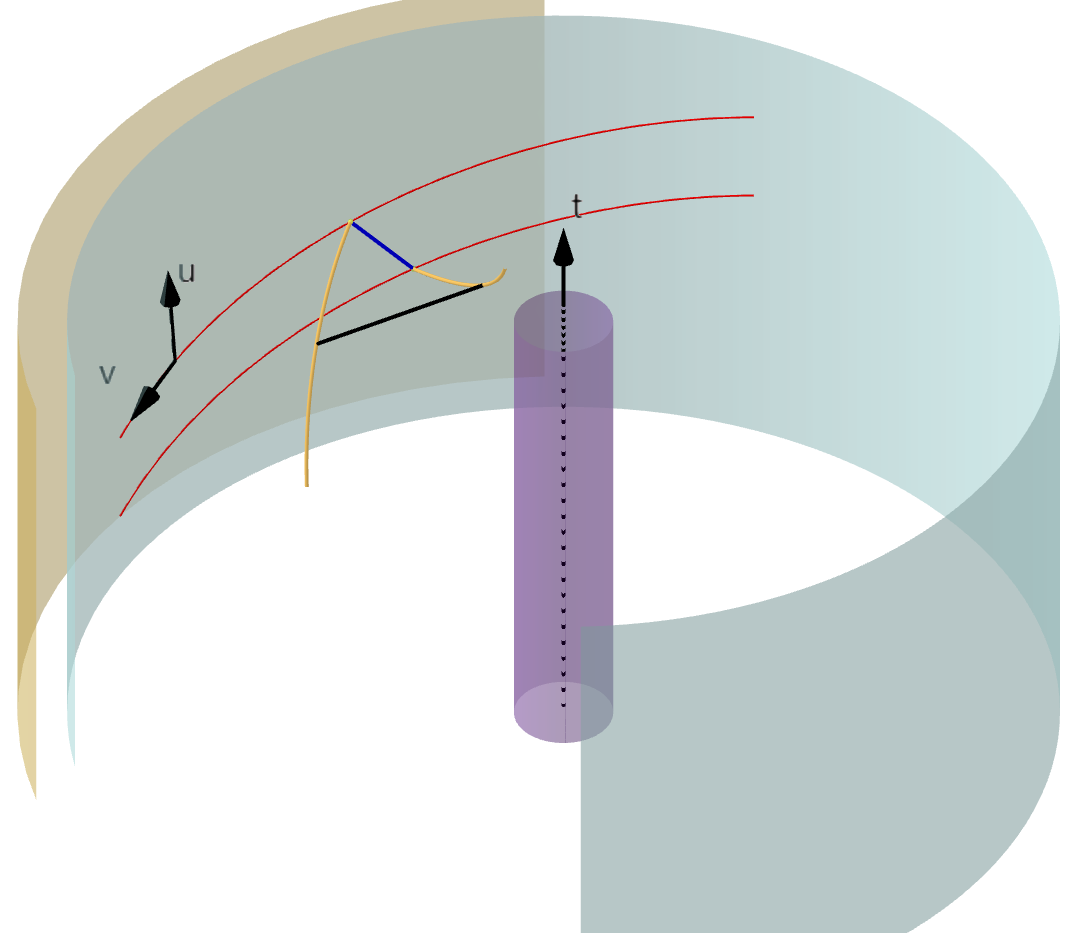}
        \caption{A cartoon picture showing the basic elements of a swing surface in the global coordinates of 3D Lorentzian BTZ black hole spacetime: the boundary interval in blue is on the (cyan) cutoff surface with $\r=\r_{\infty}$, two red lines are the boundary of its domain of dependence,  the null ropes are in yellow and the bench is in black. The yellow surface is the true spacelike conformal boundary with $\r=\infty$, the purple surface denotes the black hole horizon located at constant radius $\rho=2 T_u T_v $. The  coordinate $t$  in the Boyer-Lindquist-like coordinates $(t,r,\phi)$ is explicitly shown in the picture, and the $\phi$ direction is anticlockwise when viewing from the above.} 
        \label{fig:adswfct12}
    \end{figure}

Finally, the parametrization equations for the bench, which is an extremal (not maximal or minimal) spacelike geodesic lying between the two null ropes, are given by  
\be 
\gamma : u=\frac{u_+ + u_-}{2},\quad v\in \bigg[  \frac{v_+ + v_- -\Delta v}{2},\frac{v_+ + v_- +\Delta v}{2} \bigg] , \quad \rho= 2T_u T_v \coth{(T_u l_u)},
\ee
where the value of $\Delta v$ is
\be \label{deltav}
   \Delta v \equiv l_v +\frac{1}{T_v} \log \bigg[ \frac{\sinh{T_u l_u}}{2T_u T_v \rho_{\infty}^{-1}} \bigg].
\ee 
Thus the single interval holographic entanglement entropy on the BTZ black hole background \eqref{swing-prop} is given by 
\be \label{wcft-hee} S_{A}=\frac{T_{v}l_{v}}{4G}+\frac{1}{4G}\log{[\frac{\sinh{(T_{u}l_{u})}}{2T_{u}T_{v}\rho_{\infty}^{-1}}]} ,\ee
which matches the entanglement entropy formula \eqref{thermalent} on the thermal cylinder by the holographic map \eqref{bbmap22}, \eqref{match12} and the relations from \cite{Song:2016gtd} 
\be \b=\frac{\pi}{T_{u}}, \quad \bar{\b}=2 \pi \mu \left( \frac{T_v}{T_u}+1 \right), \quad \mu=\sqrt{\frac{-1}{4G k}}, \quad c=\frac{3}{2G}, \quad \r_{\infty}=\frac{2 T_{v}}{\e} \label{finalm11}.  \ee 

Compared with the case in AdS/CFT, there are several unusual features about the holographic entanglement entropy in  AdS$_3$/WCFT:
\begin{itemize}
 \item \textbf{Causality and unitarity.}
For a Lorentzian invariant 2D local CFT, causality is a key property which can put constraints on the behavior of entanglement entropy. For the subregion $\cA$ on a boundary Cauchy surface $\S \in \cB $, whose future and past domain of dependence $D^{+}[\S]$, $D^{-}[\S]$ together make up the whole spacetime $\cB$, there is a reduced density matrix $\r_{\cA}$ whose entanglement spectrum depends only on the domain of dependence $D[\cA]=D^+[\cA]\cup D^-[\cA]$ of the subregion $\cA$. The whole spacetime can be decomposed into four regions
 \be \cB=D^{+}[\S]\cup D^{-}[\S]=D[\cA]\cup D[\cA^{c}]\cup J^{+}[\p \cA] \cup J^{-}[\p \cA]  \label{cftbc1} \ee 
where $J^{+}(p)$ and $J^{-}(p)$ denote the boundary causal future and causal past of the point $p$. The Hilbert space $\cH$ can be decomposed into $\cH_{\cA} \otimes \cH_{\cA^{c}}$. If we unitarily evolve the reduced density matrix $\r_{\cA}$ by the transformations that are solely defined on $\cH_{\cA}$,  then the eigenvalues of $\r_{\cA}$ are not affected and so is the entanglement entropy. 

For a 2D WCFT, the background geometry is not Riemannian geometry and there is no notion about causal curves. Through the Rindler method we may identify the domain of dependence of an interval as an infinite long strip along $v$ direction, and instead of \eqref{cftbc1} we now have
\be \cB= D[\cA]\cup D[\cA^{c}] \ee 
in 2D WCFT.  In addition, in 2D WCFT the entanglement spectrum does not only depend on the domain of dependence of subregion, and different subregions having the same domain of dependence would have different entanglement entropies.  Note that in the case of 2D WCFT, the modular flow is not equal to zero at the endpoints of $\cA$, which means that we can construct an evolution operator from the modular Hamiltonian related to the subregion $\cA$ to change the length of interval in $v$ direction and thus to change the entanglement entropy.

\item \textbf{Holographic UR and IR properties:} A remarkable feature of the AdS/CFT holography is that it geometrize the energy scale of boundary field theory by bulk radial direction. From holographic entanglement entropy (HEE) point of view, the UV divergence in the area law of EE is related geometrically to the local near-boundary behavior of RT/HRT surface ending normally on the asymptotic AdS boundary, while the IR volume law contribution to EE at a finite temperature corresponds holographically  to the straddle behavior of HRT surface lying close to the horizon of black hole which is deep in the interior. For both UV and IR properties of EE in 2D WCFT, we only have similar behaviors  in the $u$ direction,  while in the $v$ direction the relation is always linear due to the non-local property of WCFT. Thus we focus only on the $u$ part of HEE in the following discussions. The swing surface touches the boundary through null ropes which have vanishing contributions to the HEE, thus the divergent UV behavior of HEE in AdS$_3$/WCFT shows up geometrically by pushing the whole bench toward to asymptotic boundary, which is very different from the way in the AdS/CFT case. While the IR property of HEE in AdS$_3$/WCFT case is behaved geometrically in the similar way as the AdS/CFT case by mainly extending the bench along the black hole horizon. 

\item \textbf{Applicability:} The holographic formula \eqref{wcft-hee} can be applied only to the zero mode background of asymptotically AdS$_3$ spacetime with parameter range $T^2_{v} \ge 0$, i.e., the BTZ black holes with non-zero charges. When $T_{v}$ has an imaginary value, it would cause troubles in several aspects. First, in \eqref{bbmap22} ,\eqref{finalm11} and \eqref{wcft-hee} the meaning for imaginary $T_{v}$ is unclear. Second, from \eqref{nullropes} we can see there would be no solutions\footnote{When $T_{u}^2=T_{v}^2=0$, there is still no sensible solution of the null ropes because this requires $ u'=0,$ $\rho'=0,$ $v'=\frac{2p_{u}}{\rho}$ whose solutions can not reach the asymptotic boundary.} of null ropes. This means that we cannot use this formula to the global AdS$_3$ spacetime and the BTZ black hole with $M=J=0$.
    
\end{itemize}


\begin{figure}
    \centering
    \includegraphics[width=0.7\textwidth]{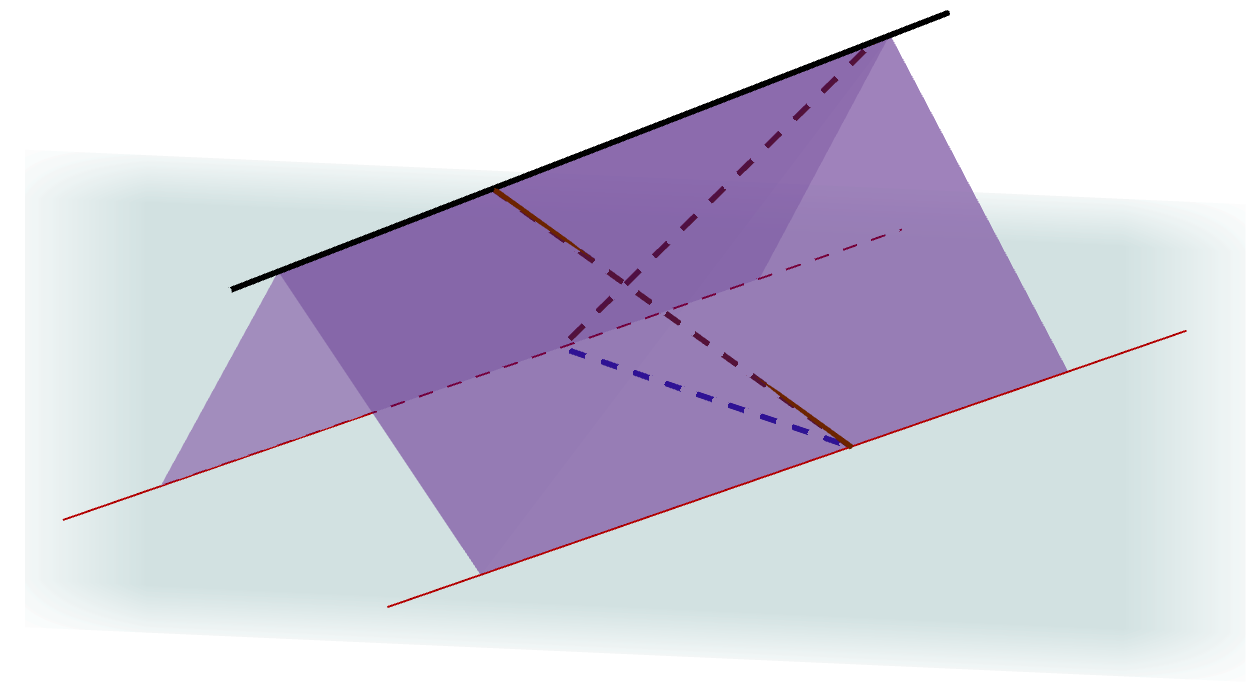}
    \caption{The figure shows schematically the infinite long pre-entanglement wedge $\cW_{\xi}$ of (blue) single boundary interval bounded by two (purple) null sheets $N_{\pm}$ and boundary domain of dependence (bounded by two red lines) in planar BTZ spacetime. For clarity we do not show explicitly the horizon here. Note that the whole $\cW_{\xi}$ lies completely outside the black hole horizon.}
    \label{fig:EWwcft}
\end{figure}

Let us explore the  entanglement wedge $\cW_{\cE}[\cA]$ of single boundary interval in AdS$_3$/WCFT. 
 In AdS$_3$/WCFT,  the bulk spacetime has well defined causal structure but the boundary field theory does not. This would lead to a tension between the cutoff surface at $\r=\r_{\infty}$ and the asymptotic boundary at $\r=\infty$, which we can see from the point of view of entanglement entropy (EE). As can be seen easily from Figure \ref{fig:adswfct12} which is consistent with bulk to boundary state-dependent map \eqref{bbmap22}, the boundary interval can be a timelike line on the cutoff surface that has well defined induced metric. In addition, it is difficult to specify a homology surface $\cR_{\cA}$ interpolating between the boundary interval $\cA$ and the bench $\g$, i.e., $ \p \cR_{\cA}=\cA \cup \g $, such that $\cR_{\cA}$ is a subregion of the bulk Cauchy surface of the whole spacetime. So the definition of entanglement wedge described in Section 2.4 can not be applied here directly. 
 
 However, there is another equivalent definition of $\cW_{\cE}[\cA]$ in AdS/CFT: it is the region containing the set of bulk points that are spacelike-separated from HRT surface $\g$ and connected to boundary domain $D[\cA]$. This definition has the advantage of freeing us from having to specify an homology surface $\cR_{\cA}$, and it depends only on  $\g$ and $D[\cA]$. We would apply this definition to AdS$_3$/WCFT. In this case, it requires starting from the bulk swing surface to grow the null normal congruence with non-positive expansion. The collection of such null normal geodesics span $N_{\pm}$, which are the future and past directed part of the light-sheets. There are potentially two candidates from which we can grow null surface, one is  the whole modular flow invariant bifurcating horizon $\g_{\xi}$ and the other one is the finite bench $\g$ that is a subregion of $\g_{\xi}$. The reason to choose the infinite long $\g_{\xi}$ is that we hope the entanglement wedge cover the whole region where bulk modular flow sitting in. Therefore we prefer the first one, and refer to the region bounded by the resulting null sheets $N_\pm$ and the boundary of domain $\partial D[\cA]$ as the pre-entanglement wedge $\cW_{\xi}$, see Figure \ref{fig:EWwcft}. It satisfies 
\be \cW_{\xi} \cap \cB= D[\cA] \ee 
for single boundary interval.

\begin{figure}
    \centering
    \includegraphics[width=0.6\textwidth]{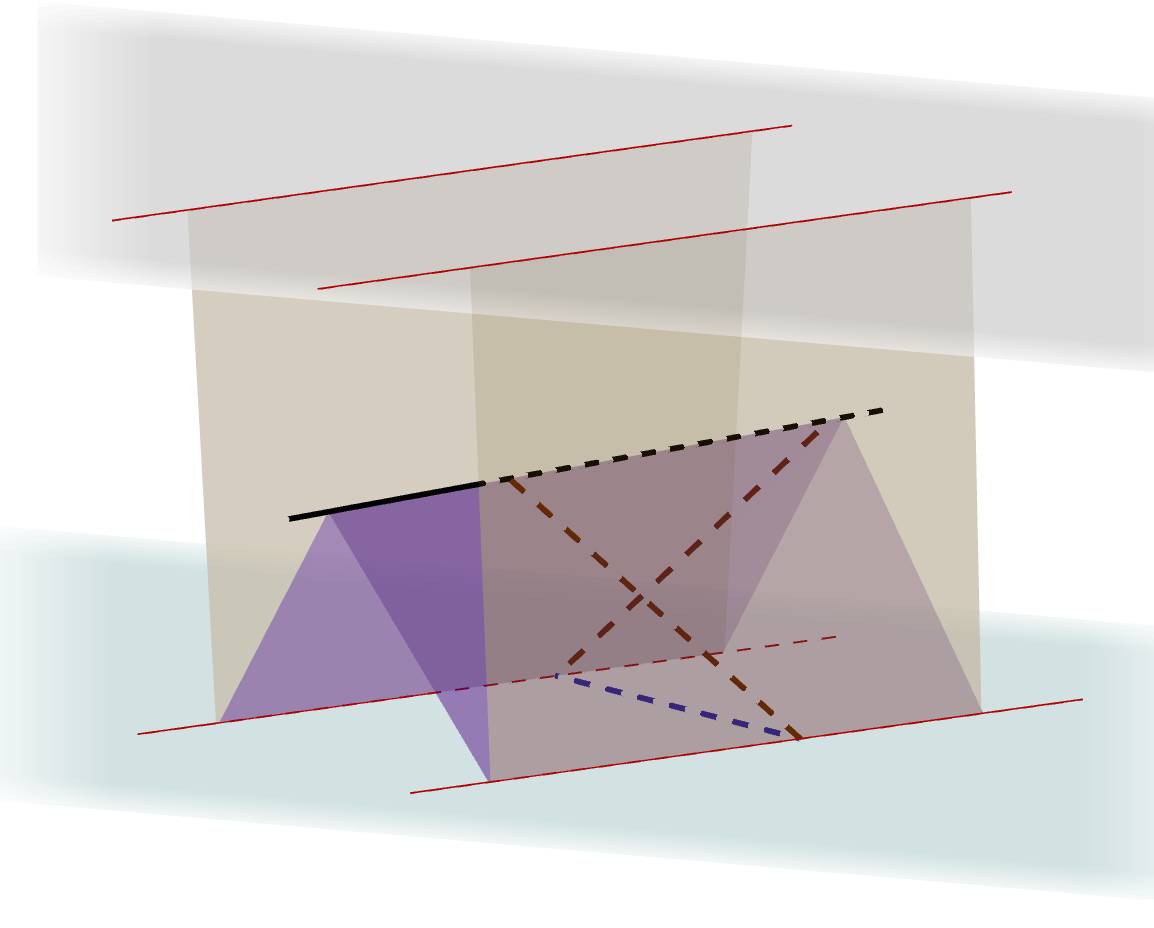}
    \caption{The relative location of the pre-entanglement wedge $\cW_{\cE}[\cA]$ (bounded by two purple surfaces) and the causal wedge $\cW_{\cC}[\cA]$ (bounded by two yellow surfaces) for (blue) single boundary interval in planar BTZ spacetime. Two red lines on the (gray) black hole horizon are the intersection of the horizon and the yellow surfaces. Obviously the causal wedge $\cW_{\cC}[\cA]$ contains the pre-entanglement wedge $\cW_{\cE}[\cA]$ in AdS$_3$/WCFT.}
    \label{fig:causalwcft}
\end{figure}
Let us turn to the causal wedge in AdS/WCFT. The definition of causal wedge in AdS/CFT is 
\be \cW_{\cC}[\cA]= \Tilde{J}^{+}[D[\cA]] \cap \Tilde{J}^{-}[D[\cA]]= \Tilde{J}^{+}[P_{1}] \cap \Tilde{J}^{-}[P_{2}],\ee
where $\Tilde{J}^{\pm}[p]$ denote the bulk causal future/past of a point $p$, and $P_{1,2}$ are the timelike future and past tips of $D[\cA]$. In WCFT the boundary domain of dependence $D[\cA]$ of interval $\cA=AB$ is an infinite long strip, which can arise from a limiting procedure defined by 
\be \lim_{P_{u_{+}},P_{u_{-}}} \Tilde{J}^{+}[P_{u_{+}}]\cap \Tilde{J}^{-}[P_{u_{-}}]\cap \cB = D[\cA] \ee  
where $\lim_{P_{u_{+}},P_{u_{-}}}$ denote the process of moving the two points $P_{u_{+}}$ and $P_{u_{-}}$, which live on the upper boundary and lower boundary of $D[\cA]$ respectively, along the $v$ direction to the future and past null infinity respectively. Then the bulk causal wedge is 
\be \cW_{\cC}[\cA]= \lim_{P_{u_{+}},P_{u_{-}}} \Tilde{J}^{+}[P_{u_{+}}] \cap \Tilde{J}^{-}[P_{u_{-}}]. \ee 
Roughly speaking, under this limiting process, the causal wedge of $\cA$ would cover as much points as it can. 
As shown in Figure \ref{fig:causalwcft}, the causal wedge contains the whole pre-entanglement wedge in AdS/WCFT, in contrast with  the  situation in AdS/CFT.


\subsection{Pre-entanglement wedge cross section and two interval entanglement phase transition}

\begin{figure}
    \centering
    \includegraphics[width=0.5\textwidth]{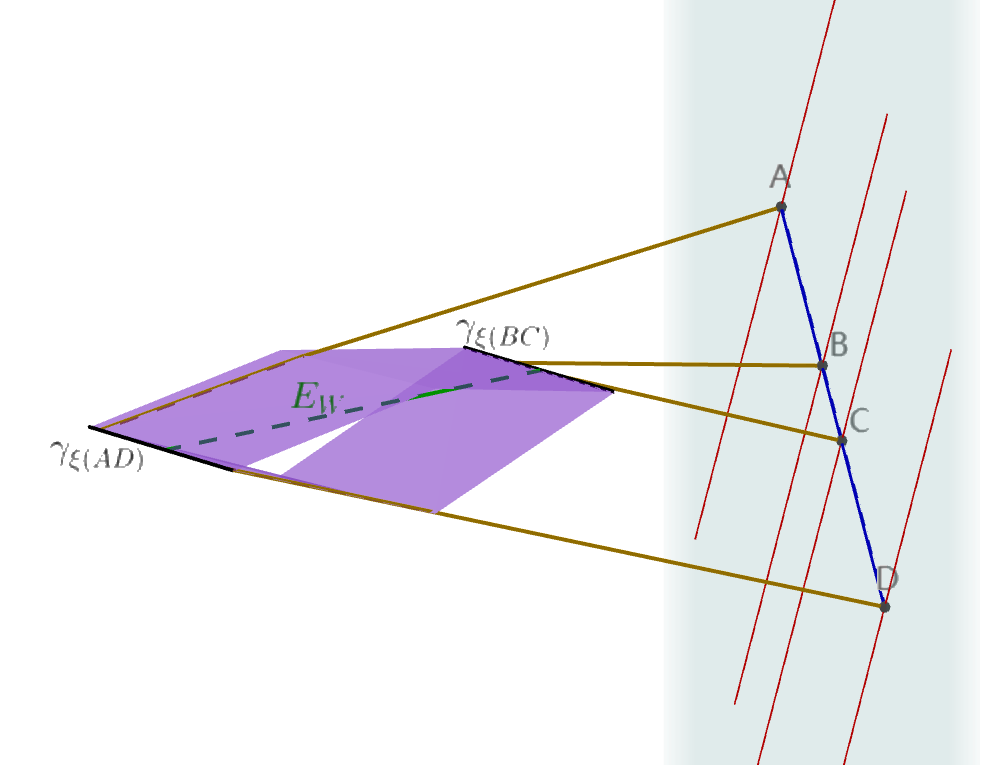}
    \caption{The figure shows the non-trivial connected pre-entanglement wedge $\cW_{\xi}[\cA]$  and its cross section $E_{W}$ (green) related to two disconnected (blue) boundary intervals $AB$ and $CD$. It is drawn in the planar BTZ black hole spacetime. For clarity,  we choose the two intervals sitting on the same line and omit the black hole horizon. $E_{W}$ has a $U(1)$ translational symmetry along the $v$ direction, which is parallel to the infinite bifurcation horizon $\g_{\xi(BC)}$ and $\g_{\xi(AD)}$. $\cW_{\xi}[\cA]$ would never intersect the asymptotic boundary, which is a new phenomena in AdS$_3$/WCFT.}
    \label{fig:EWNwcft}
\end{figure}

In this subsection, we would like to consider the pre-entanglement wedge related to two disjoint intervals in the AdS$_3$/WCFT holography.  Generally, there are two different ways of pairing the endpoints which are both consistent with the homology constraint in computing the holographic entanglement entropy of two disconnected intervals. Among them we should choose the configuration with minimal swing surface area and the corresponding entanglement wedge would follow. Like the case in AdS/CFT, an entanglement phase transition would happen in AdS$_3$/WCFT case as we change the separation of two intervals while keeping  their sizes fixed. For convenience, we choose the symmetric configuration of two boundary intervals\footnote{The result also holds for general non-symmetric configurations.}, as shown in Figure \ref{fig:EWNwcft}: 
\be  \p \cA_{1}=\{A=(-b,-c),B=(-a,-d)\}, \quad \p \cA_{2}=\{C=(a,d),D=(b,c)\}, \label{wcft-interval-ab}
\ee
with $b>a>0,c>d>0$. If the swing surface configuration pairs $A$ to $B$ and $C$ to $D$, then the area is given by
\be
    S_{\cA_1\cA_2}^{(1)}=\frac{T_v(c-d)}{2G}+\frac{1}{2G}\log\Big[\frac{\sinh(T_u(b-a))}{2T_uT_v\rho_\infty^{-1}}\Big].
\ee
If the swing surface configuration pairs $A$ to $D$ and $B$ to $C$, then the area is given by
\be
    S_{\cA_1\cA_2}^{(2)}=\frac{T_v(c+d)}{2G}+\frac{1}{2G}\log\Big[\frac{\sqrt{\sinh(2T_ub)\sinh(2T_ua)}}{2T_uT_v\rho_\infty^{-1}}\Big].
\ee
The difference of the area between these two configurations is given by
\begin{align}
     \delta S_{\cA_1\cA_2}&\equiv S_{\cA_1\cA_2}^{(2)}-S_{\cA_1\cA_2}^{(1)}=\frac{T_vd}{G}+\frac{1}{2G}\log\Big[\frac{\sqrt{\sinh(2T_ub)\sinh(2T_ua)}}{\sinh(T_u(b-a))}\Big] \nn \\
     &=\frac{T_vd}{G}+\frac{1}{4G}\log\frac{1-w}{w}.
\end{align}
where $w$ denotes the cross ratio on the thermal cylinder. This suggests that for fixed $d$, if $w\ll1$ we would have $\delta S_{\cA_1\cA_2}>0$, and we need to choose the first paring pattern. In this case, the pre-entanglement wedge $\cW_{\xi}[\cA]$ determined by the swing surface is disconnected and there is no non-trivial pre-entanglement wedge cross section. For the other limit where $1-w\ll1$ with $d$ fixed, we have $\delta S_{AB}<0$ and we should choose the second paring pattern. In this case, the corresponding pre-entanglement wedge is connected, which supports a nontrivial cross section. In addition, there always exists a non-unique spacelike homology surface $\cR_{\cA}$ interpolating between the two infinite ``benches", which gives the identification of connected pre-entanglement wedge in this situation as the bulk domain of dependence of $\cR_{\cA}$
\be \cW_{\xi}[\cA]=D[\cR_{\cA}] .\ee 

In order to compute the cross section of the connected entanglement wedge $\cW_{\xi}[\cA]$, we choose the symmetric boundary intervals satisfying \eqref{wcft-interval-ab} with the cross ratio $1-w\ll1$. The line with extremal length can only exist between two spacelike benches, $\g_{AD}$ and $\gamma_{BC}$. 
To begin with, we write down the general expression of geodesic distance $L(p_1, p_2)$ in BTZ black hole spacetime \eqref{wcft-gauge} between two spacelike separated points $p_1=(\rho_1,u_1,v_1)$ and $p_2=(\rho_2,u_2,v_2)$ for later convenience
\be 
L(p_1,p_2) =\cosh^{-1}\left( D(p_1,p_2) \right) ,\nn
\ee 
with 
\begin{align}
    & D(p_1,p_2)=D_{+}(p_1,p_2)+D_{-}(p_1,p_2), \\
    & D_{\pm}(p_1,p_2) =\frac{e^{\pm T_v v_{12}}}{8 T_u T_v} \bigg[ e^{\pm T_u u_{12} \sqrt{(\rho_1 +2T_u T_v)(\rho_2 +2T_u T_v)}} -e^{\mp T_u u_{12} \sqrt{(\rho_1 -2T_u T_v)(\rho_2 -2T_u T_v)}} \bigg] .
\end{align}
The two independent functions $D_{+}(p_1,p_2)$ and $D_{-}(p_1,p_2)$, which are invariant under $SL(2,R)_{L}$ $\times U(1)_{R}$ transformations, together constitute a quantity $D(p_1,p_2)$ that is invariant under the transformations of the group $SL(2,R)_{L}$ $\times SL(2,R)_{R}$,  the isometry group of AdS$_3$. The final form of the $L(p_1,p_2)$ can be determined from the normalization condition\cite{Apolo:2020bld}.

The parametrizations of two benches are respectively
\begin{align} 
   & \g_a : u_a=0,\quad v_a \in \bigg[  \frac{ -\Delta v_a}{2},\frac{ \Delta v_a}{2} \bigg] , \quad \rho_a = 2T_u T_v \coth(2T_u a),   \\
  & \g_b : u_b=0,\quad v_b \in \bigg[  \frac{ -\Delta v_b}{2},\frac{ \Delta v_b}{2} \bigg] , \quad \rho_b = 2T_u T_v \coth(2T_u b),
\end{align}
where 
\bea 
\Delta v_a&=&2d+\frac{1}{T_{v}}\log\left(\frac{\sinh (2 T_u a)}{2 T_u T_v \rho_{\infty}^{-1}}\right),\nn\\
\Delta v_b&=&2c+\frac{1}{T_{v}}\log\left(\frac{\sinh (2 T_u b)}{2 T_u T_v \rho_{\infty}^{-1}}\right).\nn
\eea 
We take the point $p_1$ on $\g_a$ and the point $p_2$ on $\g_b$, and find the geodesic distance between them 
\begin{align}
     D(p_1,p_2)&=D_{+}(p_1,p_2)+D_{-}(p_1,p_2) \nn \\
      &=D_{+}(u_{12}=0,v_{12},\rho_{12}=const.)+D_{-}(u_{12}=0,v_{12},\rho_{12}=const.) \nn \\
     &=\frac{A_0}{2} \cosh(T_v v_{12}).
\end{align}
where 
\be 
A_0=\sqrt{(\coth(2T_u a)+1)(\coth(2T_u b)+1)} -\sqrt{(\coth(2T_u a)-1)(\coth(2T_u b)-1)}.
\ee 
It is easy to see that there is only one minimal value of  $D(p_1,p_2)$ which corresponds to $v_{12}=0$ 
\be 
    D_{min}(p_1,p_2)=\frac{A_0}{2}=\left| \frac{\sinh(T_u (a+b))}{\sqrt{\sinh(2T_u a)\sinh(2T_u b)}} \right|.
\ee

Because the distance $L(p_1,p_2)$  is a monotonically increasing function of $D(p_1,p_2)$, then the minimal value of $D(p_1,p_2)$ directly gives the minimal value of $L(p_1,p_2)$. Using the following identities
\be 
    \cosh^{-1}(x)=\log(x+\sqrt{x^2+1}), \quad \sinh(b+a)^2-\sinh(b-a)^2=\sinh{2a}\sinh{2b},
\ee
we find the minimal length between $\g_a$ and $\g_b$
\begin{align}
     L_{min}(p_1,p_2)&=\cosh^{-1}(D_{min}(p_1,p_2))=\log \big( D_{min}+\sqrt{D_{min}^2-1}\big) \nn \\
            &=\frac{1}{2} \log \left( \left|\coth{T_u a} \tanh{T_u b}\right| \right).
\end{align}
In conclusion, we get the extremal pre-entanglement wedge cross section 
\be 
    E_W(\cA)=\frac{L_{min}(p_1,p_2)}{4G}=\frac{1}{8G}\log \left( \left| \coth( T_ua) \tanh (T_ub) \right| \right).
\ee
Comparing it with the reflected entropy of WCFT \eqref{cft-wcft-thermal}, we find that under the holographic dictionary $c=\frac{3}{2G}$ in \eqref{finalm11}, they satisfy
\be
    E_W(\cA)=\frac{1}{2}S^{\text{WCFT}}_{R;thermal},
\ee
which is precisely the conjectured holographic relation \eqref{dictionary} in AdS/CFT case. Moreover, if we take $T_u \to 0$ in \eqref{finalm11} which corresponds to $\beta \to \infty$, we would have
\be D_{min}\rightarrow \frac{1}{2} \left(\sqrt{\frac{a}{b}} + \sqrt{\frac{b}{a}}  \right) \geq 1, \qquad  L_{min} \to \frac{1}{2} \log(\frac{b}{a}). \label{tuequal0} \ee 
Then the pre-EWCS in $T_{u} =0$ zero mode background is half of  the zero temperature reflected entropy $S^{\text{WCFT}}_{R;vac}$ on the vacuum state of WCFT. 

\begin{figure}
    \centering
    \includegraphics[width=0.5\textwidth]{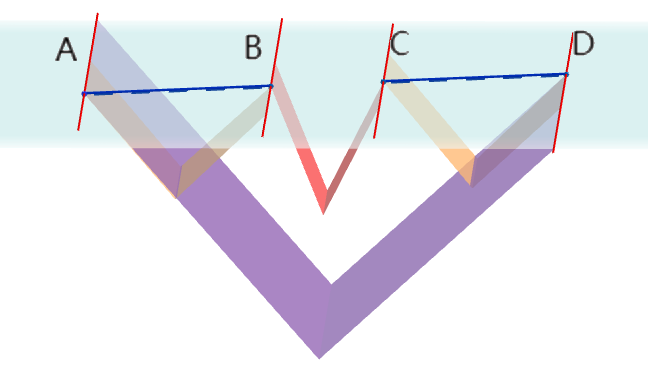}
    \caption{This cartoon picture shows the pre-entanglement wedge nesting property in AdS$_3$/WCFT holography. For two (blue) disjoint boundary intervals $AB$ and $CD$, the region bounded by the yellow null surfaces and (cyan) cutoff surface corresponds to the disconnected $\cW_{\cE}[\cA]$. The pre-entanglement wedge related to single interval $BC$ are bounded by red null surfaces, and the one related to interval $AD$ are bounded by purple null surfaces.}
    \label{fig:wcftphase3}
\end{figure}

In AdS/CFT, entanglement wedge $\cW_{\cE}[\cA]$ satisfies the entanglement wedge nesting property, which mainly states that if the two boundary subregions have the relation $\cA' \subset \cA$ then the entanglement wedge of $\cA'$ should lie within the entanglement wedge of $\cA$. It is consistent with subregion duality and implies that the sub-algebra of operators localized in $\cW_{\cE}[\cA']$ must be a subset of the sub-algebra of operators localized in $\cW_{\cE}[\cA]$. In AdS$_3$/WCFT, we still have this nice property for pre-entanglement wedge as can be seen clearly from Figure \ref{fig:wcftphase3}. The domains of dependence for the sub-regions $AB$, $BC$ and $CD$ are disconnected, and their corresponding pre-entanglement wedges are also disconnected. Furthermore, the boundary domains of dependence for the subregions $AB$, $BC$ and $CD$ are all contained in the domain of dependence of $AD$,  and the same happens for their corresponding pre-entanglement wedges. 

\section{Logarithmic negativity and odd entropy}

Additional information quantities measuring different entanglement structures of mixed state $\r_{AB}$ in many body systems have also been shown to be holographically related to the entanglement wedge cross section in  AdS$_3$/CFT$_2$. They include the logarithmic negativity $\cE(\r_{AB})$ in \cite{Kusuki:2019zsp} which is shown to be a suitable \cite{Horodecki:1996nc} and tractable measure of entanglement for mixed states, and the odd entropy which was recently introduced in \cite{Tamaoka:2018ned} as a new information quantity. It is an interesting question to ask how these two entanglement measures are related to the entanglement wedge cross section in the AdS$_3$/WCFT correspondence. 

\subsection{Logarithmic negativity}

The (logarithmic) negativity was proposed as a suitable measure of quantum entanglement for mixed states\cite{Peres:1996dw}. It is derived from the positive partial transpose  criterion for the separability of mixed states\cite{Peres:1996dw,Horodecki:1996nc,Eisert:1998pz,Simon:1999lfr}. Given  a bipartite density matrix $\rho_{AB}$, its partial transpose $\rho_{AB}^{T_B}$ is defined in terms of its matrix elements as
\begin{equation}
    \langle i_A,j_B|\rho_{AB}^{T_B}|k_A,l_B\rangle=\langle i_A,l_B|\rho_{AB}|k_A,j_B\rangle,
\end{equation}
where $i_A , j_B , k_A$ and $l_B$ are the bases for subsystems $A$ and $B$. The logarithmic negativity is then defined as
\begin{equation}
    \mathcal{E}(\rho_{AB})\equiv \log\Big|\rho_{AB}^{T_B}\Big|_1 ,
\end{equation}
where $|\mathcal{O}|_1=\mathrm{Tr}\sqrt{\mathcal{O}\mathcal{O}^\dagger}$.
When the subregion  is spherically symmetric  and the state is pure, the negativity $\mathcal{E}$ is simply the R\'enyi entropy with R\'enyi index 1/2. In this special case,  people has shown that  it satisfies\cite{Hung:2011nu,Rangamani:2014ywa}
\begin{equation}\label{single-neg}
    \mathcal{E}=\mathcal{X}_d S,
\end{equation}
where $S$ is the entanglement entropy and $\mathcal{X}_d$ is a constant which depends on the dimension of
the CFT. For 2D CFTs, its value is $\mathcal{X}_2=3/2$.

It was conjectured in \cite{Kudler-Flam:2018qjo} and proved in \cite{Kusuki:2019zsp} that the logarithmic negativity $\cE(\r_{AB})$ for two disjoint intervals in the vacuum of holographic CFT is dual to a back-reacted entanglement wedge cross section. The key point in the proof is the relation between negativity and $n=\frac{1}{2}$ Renyi reflected entropy in holographic theories, 
\be \mathcal{E}=\frac{S_{R}^{1/2}}{2} \label{neg1} .\ee 
Generally the back-reacted entanglement wedge cross section is hard to evaluate, however when considering spherically symmetric subregion configurations in the vacuum state in two dimensions we have 
\be  \mathcal{E}= \frac{3}{2} E_{W}.  \label{neg2}  \ee 
This relation is consistent with \eqref{single-neg} with $d=2$ once we note that the entanglement wedge cross section $E_W$ reduces to the entanglement entropy for a single interval. 


The logarithmic negativity of two disjoint intervals can be computed by a correlation function of twist operators
\be \mathcal{E}=\lim_{n_{e}\to 1} \log{ \langle \sigma_{n_{e}}(z_{1}) \tilde{\sigma}_{n_{e}}(z_{2}) \tilde{\sigma}_{n_{e}}(z_{3}) \sigma_{n_{e}}(z_{4})   \rangle}  ,\label{negdef} \ee 
where $n_{e}$ represent the fact that we analytically continue the even numbers of the replica sheets, and $z_{i}$ are the endpoints of these intervals. In the $t$-channel OPE expansion of four-point correlators, there appear the twist operators $\sigma_{n}^{(2)}$ and $\tilde{\sigma}_{n}^{(2)}$ which  relate the $n$-th replica sheet to the $(n\pm 2)$-th ones and have holomorphic dimensions that depend on the parity of $n$ \cite{Calabrese:2012nk},
\be h_{\sigma_{n}^{(2)}} =h_{\tilde{\sigma}_{n}^{(2)}} = \left\{
\begin{array}{rl}
\frac{c}{24}(n-\frac{1}{n}) & \text{if $n$ is odd}  ,\\
\frac{c}{12}( \frac{n}{2}-\frac{2}{n}) & \text{if $n$ is even}.   
\end{array} \right.  \label{oddeven} \ee
When we consider holographic CFTs, the Virasoro block related to $\sigma_{n}^{(2)}$ operator would be the dominant one for some specific region $x_{c}<x<1$ in the $t$-channel block expansion of the four-point correlator. We can check the above relation \eqref{neg1} \eqref{neg2} directly in symmetric interval in vacuum state
\be \mathcal{E}=\frac{S_{R}^{n=1/2}}{2}= \frac{3}{2} E_{W}=\frac{3}{4} S_{R}^{n=1}  = \frac{c}{2} \ln{\frac{1+\sqrt{z}}{1-\sqrt{z}}}. \ee


In the case of WCFT, the logarithmic negativity is determined by \eqref{negdef} as well. The twist operators appearing in the four-point correlators have the following dimensions and $U(1)$ charges. 
\begin{itemize}
    \item External twist operator $\sigma_{n_{e}}$: 
     \be  h_{\sigma_{n_{e}}}^{inv}= \frac{c}{24} \left( n_{e}-\frac{1}{n_{e}} \right) , \quad q_{\sigma_{n_{e}}}=\left( n_{e}-1 \right)\frac{i \mu k}{2}. \label{twist11} \ee 
     
    \item External twist operator $\tilde{\sigma}_{n_{e}}$: 
     \be  h_{\tilde{\sigma}_{n_{e}}}^{inv}= \frac{c}{24} \left( n_{e}-\frac{1}{n_{e}} \right) , \quad q_{\tilde{\sigma}_{n_{e}}}=-\left( n_{e}-1 \right)\frac{i \mu k}{2}. \label{twist12}  \ee 
     
     \item Internal dominant twist operator $\sigma_{n_{e}}^{(2)}$:
     \be h_{\sigma_{n_{e}}^{(2)}}^{ inv}=\frac{c}{12}\left( \frac{n_{e}}{2}-\frac{2}{n_{e}} \right), \quad q_{\sigma_{n_{e}}^{(2)}} =2( n_{e} -1)\frac{i \mu k}{2}. \ee 
     
\end{itemize}

Working in the holographic WCFTs on the vacuum state of the plane with general spectral flow parameter $\mu$, we can evaluate \eqref{negdef} assuming the WCFT block relating to the  operator $\sigma_{n_{e}}^{(2)}$ is the dominant one as in the case of reflected entropy by using \eqref{wcft4p} and \eqref{wcft 4point}. We choose the symmetric configuration 
\be A=\{(-x_{2},-y_{2}),(-x_{1},-y_{1})\}, \quad  B=\{( x_{1}, y_{1}),(x_{2},y_{2})\},  \ee 
and find 
\begin{align}
   \mathcal{E}_{WCFT} & \sim \lim_{n_{e}\to 1}  \left( \underbrace{i \sum_{j=1}^{4}q_{j}(y_{j}-\mu x_{j})+\frac{2q_{    \s_{n_{e}}  }^{2}}{k} \log{\mathrm{x}}+\frac{2q_{   \s_{n_{e}} }q_{ \tilde{\s}_{n_{e}}   }}{k} \log{(1-\mathrm{x})}}_{D_{1}=0} +\underbrace{\log{ \left(\frac{\mathcal{F}(nc-1,h_{i}^{inv},h_{\sigma_{n_{e}}^{(2)}}^{inv},\mathrm{x}) }{x_{21}^{2h_{\s_{n_{e}} }} x_{43}^{2h_{ \tilde{\s}_{n_{e}}  }}} \right)}}_{D_{2}} \right) \nn \\
   &= \lim_{n_{e}\to 1} \left( \underbrace{-\frac{q_{ \s_{n_{e}}  }^{2} }{nk} \log{\left( x_{21}x_{43} \right)}}_{D_{2,1}}+  \underbrace{  \log{ \left(\frac{\mathcal{F}(nc-1,h_{i}^{inv},h_{\sigma_{n_{e}}^{(2)}}^{inv},\mathrm{x}) }{x_{21}^{2h_{1}^{inv}} x_{43}^{2h_{3}^{inv}}} \right)} }_{D_{2,2}} \right)\nn\\
   &=\frac{1}{2}\mathcal{E}_{\text{CFT}}
\end{align}
where $q_{1,4}=q_{\sigma_{n_{e}}}$, $q_{2,3}=q_{\tilde{\sigma}_{n_{e}}}$, $h_{1,4}^{inv}=h_{\sigma_{n_{e}}}^{inv}$ and $h_{2,3}^{inv}=h_{\tilde{\sigma}_{n_{e}}}^{inv}$. In the first equality, we use \eqref{wcft4p} \eqref{wcft 4point} and the block dominance; in the second equality, $D_{1}$ part vanish due to von Neumann limit $n_{e} \to 1 $ and we use \eqref{wcftinv}; in the third equality, $D_{2,1}$ part vanish due to von Neumann limit and $D_{2,2}$ part is the same expression as the holomorphic part of negativity in holographic CFT. Thus we have proved that in AdS$_{3}$/WCFT$_{2}$, \eqref{neg1} \eqref{neg2} still holds 
\be \mathcal{E}_{WCFT}=\frac{S_{R;WCFT}^{1/2}}{2}=\frac{3}{2}E_{W}^{WCFT}=\frac{c}{4}\ln{\frac{1+\sqrt{\mathrm{x} }}{1-\sqrt{\mathrm{x}}}}. \ee 
It seems that the logarithmic negativity in holographic WCFTs is still dual to a backreacted entanglement wedge cross section. 

\subsection{Odd entropy}
Odd entanglement entropy was proposed to be another measure of mixed state\cite{Tamaoka:2018ned}.
The definition of R\'enyi version of odd entropy is 
\be  S_{o}^{n_{o}}(\rho_{AB})=\frac{1}{1-n_{o}} \left( \Tr (\rho_{AB}^{T_{B}})^{n_{o}}-1 \right),   \ee
where $T_{B}$ is the partial transposition with respect to the subsystem $B$, and $n_{o}$ represents the analytic continuation of an odd integer from which the name ``odd entropy" comes from. We can use the replica trick as in the logarithmic negativity, and evaluate it in holographic CFTs of two disjoint intervals in vacuum state, 
\begin{align}
    S_{o} &= \lim_{n_{o} \to 1} \frac{1}{1-n_{o}} \left[ \langle \sigma_{n_{o}}(z_{1}) \tilde{\sigma}_{n_{o}}(z_{2}) \tilde{\sigma}_{n_{o}}(z_{3}) \sigma_{n_{o}}(z_{4})   \rangle-1 \right] \nn\\
         & \sim -\partial_{n_{o}} \left[ b_{\sigma_{n}^{(2)}} (z_{14}z_{23})^{-4h} \mathcal{F}(nc,h_{i},h_{\sigma_{n}^{(2)}},z) \bar{\mathcal{F}}(n \bar{c},\bar{h}_{i},\bar{h}_{\sigma_{n}^{(2)}},\bar{z}) \right]_{n_{o}=1} \nn\\
         &= \frac{c}{3} \log{\frac{z_{14}}{\epsilon}}+\frac{c}{3} \log{\frac{z_{23}}{\epsilon}}+\frac{c}{6} \ln{\frac{1+\sqrt{z}}{1-\sqrt{z}}} \nn\\
         &= S_{\rho_{A}\otimes \rho_{B}}+E_{W}(\rho_{AB})
\end{align}
where we used the odd integer part of \eqref{oddeven} and $S_{\rho_{A}\otimes \rho_{B}}$ represents the entanglement entropy of the product state $\rho_{A}\otimes \rho_{B}$. The effect of OPE coefficients $b_{\sigma_{n}^{(2)}}$ can be evaluated in the pure state limit \cite{Tamaoka:2018ned}.


For WCFT, the external twist operators in computing the odd entropy are the same as the ones in computing negativity, and their dimensions and charges are given by \eqref{twist11} \eqref{twist12} with $n_e$ being replaced by $n_o$. But the internal dominant twist operators $\sigma_{n_{o}}^{(2)}$  have different dimensions and $U(1)$ charges,
\be  h_{\sigma_{n_{o}}^{(2)}}^{inv}=  h_{\sigma_{n_{o}}}^{inv}=\frac{c}{24}\left(  n_{o} -\frac{1}{n_{o}} \right), \quad q_{\sigma_{n_{o}}^{(2)}} = 2( n_{o} -1)\frac{i \mu k}{2}  \ee 
which is consistent with \eqref{oddeven}. Assuming the block dominance in the holographic WCFT, we get the odd entropy on the reference plane\footnote{The odd entropy of WCFT is not spectral flow invariant and depends on the plane on which it is evaluated.}
\begin{align}
    S_{o}^{WCFT} &=\lim_{n_{o} \to 1}{\frac{1}{1-n_o}} \left[ \langle \sigma_{n_{o}}(x_1,y_1) \tilde{\sigma}_{n_{o} }(x_2,y_2) \tilde{\sigma}_{n_{o}}(x_3,y_3) \sigma_{n_{o}} (x_4,y_4) \rangle_{{\mbox{\small WCFT}^{\otimes n_o}(ref)}}-1 \right]\nn\\
    & \sim - \partial_{n_o} \left[ \underbrace{e^{i q_{\sigma_{n_{o}} (y_{1}+y_{4}-y_{2}-y_{3})}} \left(x_{14} x_{23} \right)^{-2h_{\sigma_{n_{o}}}} }_{D_{1}}  \underbrace{b_{\sigma_{n_{o}}^{(2)}}^{2} \mathcal{F}(nc-1,h_{i}^{inv},h_{\sigma_{n_{o}}^{(2)}}^{inv},\mathrm{x} )}_{D_2}  \underbrace{\left( \frac{\mathrm{x}}{1-\mathrm{x}} \right)^{2q_{\sigma_{n_{o}}}^{2} }}_{D_{3}} \right].\nn
 \end{align}
Here the $D_3$ part has vanishing contribution and $D_2$ leads to $E_W(\rho_{AB})$, but  the $D_{1}$ part does not give the expected $S_{\rho_{A}\otimes \rho_{B}}$ which should come from a term like 
\be S_{\rho_{A}\otimes \rho_{B}}= - \partial_{n} \left[  e^{i  q_{\sigma_{n_{o}}(y_{1}-y_{4}-y_{2}+y_{3})}} \left(x_{14} x_{23} \right)^{-2h_{\sigma_{n_{o}}}} \right]_{n_o=1}.  \ee 
In short, for the holographic WCFT, we find 
\be 
S_{o}^{WCFT} \neq S_{\rho_{A}\otimes \rho_{B}}+E_W(\rho_{AB}). \label{oddine}
\ee 
This is a new phenomena in the holographic WCFT.  

\section{Conclusion and Discussion}

In this work, we studied the reflected entropy in two-dimensional holographic warped conformal field theory, and paid special attention to  its implication in holography. We showed that in the AdS$_3$/WCFT correspondence, the reflected entropy of two disjoint intervals is related to the pre-entanglement wedge cross-section (pEWCS)
\be 
S_R(A:B)=2\frac{\mbox{Area(pEWCS)}}{4G_N}, \label{RpEWCS}
\ee 
up to the order ${\cal O}(1)$ quantum correction. This provides a nontrivial check on the relation between the reflected entropy and EWCS beyond the AdS/CFT correspondence. 

The physical implications of the relation \eqref{RpEWCS}) are  remarkable. On one hand,  it supports the two-sided dynamical construction of EWCS in \cite{Dutta:2019gen}, as described in Section 2.4. On the other hand, it provides more evidence on both the AdS$_3$/WCFT correspondence and the swing surface proposal. Most of the previous studies focused on the entanglement entropy of single interval and its holographic description. In the two-interval case, one may compute the R\'enyi mutual information on the field theory side, but short of bulk computation\cite{Chen:2019xpb}. The present study on the entanglement of double intervals provides nontrivial check on the swing surface proposal of holographic entanglement entropy. 

Another remarkable point from our study is on the semi-classical gravitational action in AdS$_3$/WCFT. It is not hard to notice that both the reflected entropy and logarithmic negativity\footnote{The situation in the odd entropy is special. We have no idea why there is \eqref{oddine}.} in holographic WCFT take half of the values of the ones in holographic CFT. This does not mean that the semiclassical  action of the AdS$_3$ gravity with the $CSS$ boundary condition is half of the one of the usual AdS$_3$ gravity. The subtle point is that these two quantities are computed by taking $n\to 1$ limit, and there could be extra terms in the gravitational action  proportional to $(n-1)$, as shown in  the R\'enyi mutual information\cite{Chen:2019xpb}.

Although we accumulate some features about pre-entanglement wedge $\cW_{\xi}[\cA]$ and causal wedge $\cW_{\cC}[\cA]$ related to single interval and two disconnected intervals in the AdS$_3$/WCFT holography, the implications of these results are not very clear. We expect further explorations in this direction. 

In our study on the field theory side, we computed the dimensions and charges  of various twist operators in non-Abelian orbifold WCFT, which allow us to read the reflected entropy, the logarithmic negativity and the odd entropy in the holographic WCFT. One essential assumption in our study is the block dominance in the large $c$ limit. One interesting question is to justify this assumption. Another important question is to find the next-to-leading order correction to the reflected entropy in the large $c$ limit, which should correspond to the quantum correction to \eqref{RpEWCS}. This may require a careful study of operator product expansion of twist operators, as shown in \cite{Chen:2013kpa,Chen:2013dxa,Chen:2019xpb}.  

\section*{Acknowledgments}
We are grateful to Peng-xiang Hao for initial participation in the project. We would like to thank Wei Song and Peng-xiang Hao for many valuable discussions.  The work is in part supported by NSFC Grant  No. 11735001. 

\printbibliography

\end{document}